\documentclass[12pt]{article}
\usepackage[english]{babel}
\usepackage{color}
\usepackage{graphicx}
\usepackage{amsmath}
\usepackage{amsthm}
\usepackage{amssymb}
\usepackage[hyperindex,plainpages=false]{hyperref}
\usepackage[left=1in,right=1in,bottom=1in,top=1in,a4paper]{geometry}

\title{B\"acklund transformations for the Camassa-Holm equation}
\author{Alexander G. Rasin  \\
Department of Mathematics,\\ 
Ariel University, Ariel, 40700, Israel \\
{E-mail: rasin@ariel.ac.il}\and  Jeremy Schiff \\
Department of Mathematics,\\
Bar-Ilan University, Ramat Gan, 52900, Israel \\
{E-mail: schiff@math.biu.ac.il}}

\begin{document}
\maketitle

\begin{abstract}
The B\"acklund transformation (BT) for the Camassa-Holm (CH) equation is presented and discussed. Unlike the vast majority of BTs
studied in the past, for CH the transformation acts on both the dependent and (one of) the
independent variables. Superposition principles are given for the action of double BTs on 
the variables of the CH and the potential CH  equations. Applications of the BT and its superposition 
principles are presented, specifically the construction of travelling wave solutions, 
a new method to construct multi-soliton, multi-cuspon and soliton-cuspon solutions, and a
derivation of generating functions for the local symmetries and conservation laws of the CH hierarchy. 
\end{abstract}

\section{Introduction}

The original B\"acklund transformation (BT) arose in the context of differential geometry of surfaces 
in the 1880s \cite{Ba1}. In the modern era, BTs have been recognized as playing a central role in the 
theory of integrable differential equations \cite{Lamb,RogS1,RogS2}. 
Their primary application is as a method to generate explicit solutions, exploiting the so-called superposition principle,
an algebraic rule to ``combine'' two solutions obtained by BTs (from a given initial solution). However, in recent work
\cite{RS2} we have also shown how to derive local symmetries and conservation laws directly from BTs. There is also 
a deep relationship between BTs and the associated linear systems of integrable equations. 

The Camassa-Holm (CH) equation \cite{CH0,CHH} is by now recognized as one of the archetypes
of integrable equations. It has (weak) ``peakon'' solutions --- solitary waves with discontinuous first derivative at their 
crest ---  and numerous other types of travelling wave solution, including solitons (smooth solitary
waves), cuspons and various periodic structures \cite{LO1,LO2,Boyd1,Boyd2,KZ1,Park1,Park2,PARKER1,Len1,PV0}. 
The integrability of the CH equation was already firmly established in \cite{CH0}, where 
a Lax pair and a bihamiltonian structure were given, and much further evidence for this has
accumulated since then. There is an inverse scattering formalism \cite{cons1,cons2},
explicit formulas can be found for multipeakon,multisoliton, multicuspon and soliton-cuspon solutions 
\cite{Je1,BSS1,BSS4,BSS2,KZ2,Johnson,LZ1,Park3,Dai1,Li,MAT2,PM1,MAT1,PARKER2,PARKER3,Dai3,XZQ} 
there are an infinite number of local conservation laws \cite{FS1,rey0,rey1,HR1,reyP,HR2,Len2,CLOP,Iv1,GKKV}, 
and there is a rich algebra of symmetries \cite{rey0,rey1,HR1,reyP,HR2,GKKV}.
Other significant works on CH include studies of the stability of peakon 
and other exact solutions \cite{cons3,cons4,cons5,Len5} and interesting numerical studies 
\cite{Len3,MFO2,MFO1,CLP}.

The aim of this paper is to fully explore the theory of the BT for the CH equation. 
In \cite{Je1}, one of us constructed a BT for the associated CH (aCH) equation, an equation related by a (field dependent)
change of coordinates to the CH equation, and used this to construct some solutions of CH which could be 
regarded as superpositions of 2 travelling waves. However, this work was incomplete; 
an integration was required to reconstruct a solution of CH from a solution of aCH, 
which, in general, could not be done explicitly, severly limiting applicability. 
In the current paper we resolve this and other problems. The BT of CH differs from 
standard ones (for example, those of KdV and  Sine-Gordon) 
in that {\em it involves a transformation of both the dependent and one of the independent variables}. 
However, remarkably, there is a nonlinear superposition principle for both of these transformations, 
which we develop and apply to the generation of multisoliton, multicuspon and soliton-cuspon solutions, 
as well as to the derivation of symmetries and conservation laws for CH.  
The action of the BT on both dependent and independent variables is not unique to CH;
a similar situation exists for the Dym equation, which also exhibits nonanalytic solitons  
\cite{ty1,SM1}.

The structure of this paper is as follows: In section 2 we recap the known results for the aCH equation. 
In section 3 we use them to derive the BT for CH. Section 4 discusses the various forms of superposition principle. 
In section 5 we use the BT to obtain travelling wave solutions. The BT is used to construct soliton and cuspon 
solutions from which the standard peakon solutions can be obtained in a certain limit. Alas it does not seem to 
give a direct construction of peakons. However, various other unphysical solutions are also obtained.   
In section 6 we use the superposition principle to obtain 
cuspon-cuspon, soliton-soliton and cuspon-soliton solutions. 
In section 7, following \cite{RS2},  we use the BT to construct the conservation laws and symmetries of CH. 
Section 8 contains some concluding remarks. 

\section{Previous results}
The Camassa-Holm equation (CH) \cite{CH0} is
\begin{equation}
m_t+2u_xm+um_x=0,\qquad    m=u-u_{xx},\label{CM0}
\end{equation}
or equivalently 
\begin{equation}
u_t-u_{txx}+3uu_x-uu_{xxx}-2u_xu_{xx}=0.  \label{CM1} 
\end{equation}
By translating $u$ and performing a Galilean transformation $x \rightarrow x-ct$ it is possible to introduce
linear transport and 
linear dispersion terms into the equation, see for example \cite{DGH}. All the results we present here can be 
generalized for the full  class of equations considered in \cite{DGH}. 

Writing $u=v_x$ and integrating once, we obtain the potential Camassa-Holm equation (pCH) 
$$
v_t-v_{txx}+\frac{3}{2}v_x^2-v_xv_{xxx}-\frac{1}{2}v_{xx}^2=0 \ , 
$$
or, equivalently,  
$$
n_t  + \frac12 v_x^2 + v_xn_x - \frac12 (v-n)^2 = 0 \ , \qquad n = v - v_{xx}\ . 
$$
Evidently $n$ is a potential for $m$,  $m=n_x$. 

In \cite{Je1} equation (\ref{CM0}), under the assumption $m>0$, 
was transformed to the associated Camassa-Holm equation (aCH)
$$
2p_{\tau}=-p^2u_{\xi}\ , \qquad  u=-\frac{p}{2}\left(\frac{p_{\tau}}{p}\right)_{\xi}+p^2 \ ,
$$
with the help of transformation
\begin{equation}
p=\sqrt{m}\ ,\qquad d\xi=\frac{1}{2}pdx-\frac{1}{2}pudt\ ,\qquad   \tau=t.\label{tr1}
\end{equation}
This transformation implies
\begin{equation}
\frac{\partial x}{\partial \xi}=\frac{2}{p}\ ,
\qquad\frac{\partial x}{\partial \tau}=u\ ,
\qquad\frac{\partial t}{\partial \xi}=0\ ,
\qquad\frac{\partial t}{\partial \tau}=1 \ .\label{tr2}
\end{equation}

A BT for aCH was found in \cite{Je1}:  
\begin{equation}
p\rightarrow p-s_{\xi}\ , \qquad
u\rightarrow u+\frac{2s_{\tau}}{p(p-s_{\xi})},\label{BT10}
\end{equation}
where $s$ satisfies
\begin{align}
s_{\xi}=&-\frac{s^2}{p\alpha}+\frac{\alpha}{p}+p\ ,\label{tr3}\\
s_{\tau}=&-s^2+\frac{p_{\tau}}{p}s+\alpha(\alpha+u)\ .\label{tr4}
\end{align}
The following nonlinear superposition principle was also given: 
\begin{equation}
p\rightarrow p-\left(\frac{(\alpha-\beta)(\alpha\beta-s_{\alpha}s_{\beta})}{\beta s_{\alpha}-\alpha s_{\beta}}\right)_{\xi}
 \   ,\label{BTaCH}
\end{equation}
where $s_{\alpha}, s_{\beta}$ are the solutions of (\ref{tr3},\ref{tr4}) with parameters $\alpha$ and $\beta$ respectively.

In \cite{RS2} the BT was used to find an infinite number of symmetries for aCH. These are given by the generating symmetry 
$X=Q^p\frac{\partial}{\partial p}+Q^u\frac{\partial}{\partial u}$ where
\begin{equation}
Q^p=\frac{p(s_{\alpha}^{(1)}+s_{\alpha}^{(2)})}{\alpha(s_{\alpha}^{(1)}-s_{\alpha}^{(2)})}\ ,
\qquad 
Q^u=-\frac{2s_{\alpha}^{(1)}+2s_{\alpha}^{(2)}+pu_{\xi}}{s_{\alpha}^{(1)}-s_{\alpha}^{(2)}}.\label{QQ1}
\end{equation}
Here $s_{\alpha}^{(1)},s_{\alpha}^{(2)}$ are two different solutions of (\ref{tr3},\ref{tr4}) for the same parameter $\alpha$.
This symmetry depends upon $\alpha$; expansion in a (formal) power series in $\alpha$ 
gives the infinite hierarchy of symmetries.

\section{The B\"acklund transformation for the Camassa-Holm equation}
In this section we obtain the BT for CH and pCH from the BT for aCH. 
With the help of (\ref{tr2}) we write the BT (\ref{BT10}),(\ref{tr3}),(\ref{tr4}) as 
\begin{equation}
u\rightarrow u-2\alpha-\frac{2\alpha(u_xs-\alpha u)}{s^2-\alpha^2}\ ,\label{BT4}
\end{equation}
where $s$ satisfies
\begin{align}
s_{x}=&-\frac{s^2}{2\alpha}+\frac{1}{2}(m+\alpha)\ ,\label{BT1}\\
s_{t}=&-s^2\left(1-\frac{u}{2\alpha}\right)-u_xs+\frac{1}{2}(2\alpha^2+\alpha u-um)\ .\label{BT2}
\end{align}
This system for $s$ is equivalent to the Lax pair for CH. 
Note (\ref{BT2}) can be simplified with the help of (\ref{BT1}) and (\ref{CM0}) to 
\begin{equation}
s_t=\alpha u_{xx}+ 2\alpha s_x -u s_x -u_{x}s\ .\label{BTT}
\end{equation}
In light of (\ref{tr2}) the BT for CH must also involve the independent variable $x$.
Using the first equation in  (\ref{tr2}), the change of the independent variable is
\begin{eqnarray*}
x_{\rm new} - x   &=& \int \left( \frac{2}{p_{\rm new}}  - \frac{2}{p} \right )  \ d \xi  \\
                &=& \int \left( \frac{2}{p-s_\xi}  - \frac{2}{p} \right )  \ d \xi  \\
                &=& \int \frac{2s_\xi}{p(p-s_\xi)}  \ d \xi  \\
                &=&  \int \frac{2 ds}{\frac{s^2}{\alpha} - \alpha} \\
                &=&  \ln \left| {\frac {s-\alpha}{s+\alpha}} \right| + f(\tau) \ .  
\end{eqnarray*}
In moving from the third to the fourth line here the formula for $s_\xi$ in (\ref{tr3}) is used 
in the denominator but not in the numerator. The integration leaves undetermined 
an arbitrary function $f(\tau)$. Using the 
second equation in  (\ref{tr2}) it is straightforward to show this must be a constant, 
which can be taken, without loss of generality, to be zero. 
Thus the effect of the BT on the independent coordinates is 
\begin{equation}
x\rightarrow x+\ln  \left| {\frac {s-\alpha}{s+\alpha}} \right|, \qquad  t\rightarrow t. \label{BT3}
\end{equation}
There is no guarantee that this mapping will be a bijection. We will see later an example 
in which the BT generates several  solutions out of one, in the case that this mapping is not $1$ to $1$. 

Using (\ref{BT10}) and (\ref{tr3}) it is straightforward to write down the BT for the field $p$
\begin{equation}
p \rightarrow  \frac{s^2-\alpha^2}{\alpha p} 
\label{BTp}
\end{equation}
and hence also for the field $m=p^2=u-u_{xx}$ 
\begin{equation}
m\rightarrow \frac { \left( s^2 - {\alpha}^{2}  \right) ^{2}}{{\alpha}^{2}m}\ . 
\label{BTm}\end{equation}
Further calculations give the action of the BT  for the pCH fields
$v$ (satisfying $u=v_x$) and $n=v-v_{xx}$: 
\begin{equation}
n\rightarrow n-2s\ ,\label{nn}
\end{equation}
\begin{equation}
v\rightarrow v+\frac{2\alpha(\alpha u_{x}-us)}{s^2-\alpha^2}\ . \label{BTv1}
\end{equation}

As mentioned above, the BT can be generalized for the full family of equations from \cite{DGH} 
\begin{equation}
c_1 u_x + c_2 u_{xxx} + c_3 (u_t + 3 uu_x) = c_4 (u_{txx} + uu_{xxx} + 2 u_x u_{xx} )  \ ,  \label{GCH}
\end{equation}
where $c_1,c_2,c_3,c_4$ are constants.
(This generalized equation is referred to in \cite{GL,ZCH} as the ``CH-r equation''.) The BT is
$$
x  \rightarrow  x+\sqrt{\frac{c_4}{c_3}}\ln\left|\frac{s\sqrt{c_4}-\alpha\sqrt{c_3}}{s\sqrt{c_4}+\alpha\sqrt{c_3}}\right| \ , \qquad
u  \rightarrow  u-2\alpha-\frac{2c_3\alpha^2u - 2 c_4\alpha su_x + c_1\alpha^2 + c_2s^2}{c_3\alpha^2-c_4 s^2}  \ . 
$$
Here $s$ satisfies 
\begin{eqnarray}
s_{x}&=&-\frac{s^2}{2\alpha}+\frac{\alpha}{2} \frac{2c_3 u- 2c_4 u_{xx} + 2c_3\alpha+c_1}{2 c_4 \alpha  - c_2},\\
s_t&=&\alpha u_{xx}+2\alpha s_x-us_x-u_{x}s.
\end{eqnarray}
Equation (\ref{GCH}) includes the  KdV, CH, and Hunter-Saxton (HS) \cite{HZ} equations.
The KdV equation can be obtained by putting $c_1=c_4=0$. The HS equation 
\begin{equation}
u_{tx}+\frac{1}{2}u_x^2+uu_{xx}=0 \label{HS}
\end{equation}
can be obtained  by putting $c_1=c_2=c_3=0$ and integrating with respect to $x$. 
The BT in this case is 
$$
x \rightarrow x-\frac{2\alpha}{s}\ , \qquad 
u \rightarrow u-\frac{2\alpha u_x}{s}-2\alpha,
$$
where  
\begin{eqnarray}
s_{x}&=&-\frac{s^2}{2\alpha}-\frac{u_{xx}}{2},\label{BTHS1}\\
s_t&=&\alpha u_{xx}  + 2\alpha s_x - us_x -  u_{x}s.\label{BTHS2}
\end{eqnarray}

\section{The double B\"acklund transformation and superposition principles}
In this section we discuss double BTs for CH and pCH. 
We also show the superposition principles for these equations.

As we saw in the previous section, a BT (which acts on the CH fields $u,m,p=\sqrt{m}$, the pCH fields $v,n$ and the 
independent coordinate $x$ according to equations 
(\ref{BT4}), (\ref{BTm}), (\ref{BTp}), (\ref{BTv1}), (\ref{nn}), (\ref{BT3})
respectively) is determined by a solution $s$ of (\ref{BT1}),(\ref{BT2}). We use the following notation:
Denote by $s_\alpha$, $s_\beta$ etc the solutions of (\ref{BT1}),(\ref{BT2}) corresponding to parameters
$\alpha,\beta$ etc. Denote the associated action on the fields by $u\rightarrow u_\alpha$, $m\rightarrow m_\alpha$ 
etc. Denote by $s_{\alpha\beta}$ the solution of (\ref{BT1}),(\ref{BT2}) with $u,m$ replaced by $u_\alpha,m_\alpha$ and 
parameter $\beta$ (i.e. we start with a solution obtained from a BT with parameter $\alpha$ and are now 
considering acting upon it by a further BT with parameter $\beta$). Denote the corresponding action on the 
fields by $u_\alpha\rightarrow u_{\alpha\beta}$, $m_\alpha\rightarrow m_{\alpha\beta}$ etc. 

The fundamental fact about double BTs, as proved in \cite{Je1}, is that they commute, i.e. 
$u_{\alpha\beta}=u_{\beta\alpha}$, $m_{\alpha\beta}=m_{\beta\alpha}$ etc. From, for example, the transformation law for the pCH 
field $n$, (\ref{nn}), it immediately follows that 
\begin{equation}
s_\alpha  + s_{\alpha\beta}  = s_\beta + s_{\beta\alpha}  \ . \label{comm}
\end{equation} 
Checking the consitency of this with the versions of (\ref{BT1}) and (\ref{BT2}) satisfied by 
$s_\alpha,s_\beta,s_{\alpha\beta},s_{\beta\alpha}$  we obtain 
\begin{equation}
  s_{\alpha\beta}=-s_\alpha + \frac{(\alpha-\beta)(\alpha\beta-s_\alpha s_\beta)}{\beta s_\alpha-\alpha s_\beta} \ , \qquad 
  s_{\beta\alpha}=-s_\beta + \frac{(\alpha-\beta)(\alpha\beta-s_\alpha s_\beta)}{\beta s_\alpha-\alpha s_\beta} \ . 
\label{BTss}
\end{equation}
In fact it is possible to check directly that these formulas for $s_{\alpha\beta},s_{\beta\alpha}$ 
give solutions of the relevant versions of (\ref{BT1}) and (\ref{BT2}) without any need to assume (\ref{comm}). 

From (\ref{BTss}) it follows that once $s_\alpha$ and $s_\beta$ are known, it is possible to immediately 
find the action of a double BT. Using the transformation laws for $m,p,n,x$ and (\ref{BTss}) we find
\begin{eqnarray}
p_{\alpha\beta} &=&  \frac{\alpha\beta\left((s_\alpha-s_\beta)^2 - (\alpha-\beta)^2   \right) }{(\beta s_\alpha-\alpha s_\beta)^2}\  p   
    \label{spp}\\
m_{\alpha\beta} &=&  \frac{\alpha^2\beta^2\left((s_\alpha-s_\beta)^2 - (\alpha-\beta)^2   \right)^2 }{(\beta s_\alpha-\alpha s_\beta)^4}\  m   
     \label{spm}\\
n_{\alpha\beta} &=&  n - 2 \frac{(\alpha-\beta)(\alpha\beta-s_\alpha s_\beta)}{\beta s_\alpha-\alpha s_\beta}   
    \label{spn}\\
x_{\alpha\beta} &=&  x + \ln\left|    \frac{s_\beta-s_\alpha+\alpha-\beta}{s_\beta-s_\alpha-\alpha+\beta}   \right|  
    \label{spx}
\end{eqnarray}
For $u$ and $v$ we proceed as follows. From (\ref{BT4}) and (\ref{BTv1}) we obtain 
\begin{equation}
u_{\alpha}+u+\frac{1}{\alpha}(v_{\alpha}-v)s_{\alpha}=-2\alpha\ \label{sp1} 
\end{equation}
and similarly 
\begin{eqnarray}
u_{\beta}+u+\frac{1}{\beta}(v_{\beta}-v)s_{\beta}&=&-2\beta\ , \label{sp2} \\
u_{\alpha\beta}+u_{\beta}+\frac{1}{\alpha}(v_{\alpha\beta}-v_{\beta})s_{\beta\alpha}&=&-2\alpha\ ,\label{sp3} \\
u_{\alpha\beta}+u_{\alpha}+\frac{1}{\beta}(v_{\alpha\beta}-v_{\alpha})s_{\alpha\beta}&=&-2\beta\ . \label{sp4} 
\end{eqnarray}
Eliminating $v_\alpha,v_\beta,v_{\alpha\beta}$ from these 4 relations, using (\ref{BTss}) for $s_{\alpha\beta}$ and 
(\ref{BT4}) for $u_\alpha,u_\beta$ we obtain 
\begin{equation}
u_{\alpha\beta} =u- \frac{ 2( \alpha-\beta)\left( (\alpha-\beta)(\alpha+\beta+u) + (s_\beta-s_\alpha)(s_\beta+s_\alpha+u_x)  \right)  } 
                      { (\alpha-\beta)^2 - (s_\beta-s_\alpha)^2 }   \ . \label{spu}
\end{equation} 
Similarly, by first eliminating the fields $u$,
\begin{equation}
v_{\alpha\beta} =v-\frac{ 2( \alpha-\beta)\left( (\alpha-\beta)u_x  + (s_\beta-s_\alpha)u  + 2  (\alpha s_\beta - \beta s_\alpha)  \right)  } 
                      { (\alpha-\beta)^2 - (s_\beta-s_\alpha)^2 }   \ . \label{spv}
\end{equation} 

Equations (\ref{spp}),(\ref{spm}),(\ref{spn}),(\ref{spx}),(\ref{spu}) and (\ref{spv}) are algebraic 
formulas for the implementation of a double BT given $s_\alpha$ and $s_\beta$. However $s_\alpha$ and 
$s_\beta$ also determine the implementation of the original single BTs, so it is 
natural to try to eliminate them  
to obtain nonlinear superposition formulae for each of the quantities $p,m,n,x,u,v$. For example, for $x$
we have, from (\ref{BT3}),  
$$ s_\alpha = \frac{e^x+e^{x_\alpha}}{e^x-e^{x_\alpha}}\ \alpha \ , \qquad 
   s_\beta = \frac{e^x+e^{x_\beta}}{e^x-e^{x_\beta}}\ \beta 
$$ 
and using these in (\ref{spx}) gives 
\begin{equation}
\frac{(e^x-e^{x_\alpha})(e^{x_\beta}-e^{x_{\alpha\beta}})}{(e^x-e^{x_\beta})(e^{x_\alpha}-e^{x_{\alpha\beta}})}
  = \frac\alpha\beta\ . 
\end{equation} 
Thus we see {\em $e^x$ satisfies the cross-ratio equation}, equation A1[$\delta=0$] in the ABS classifciation
\cite{ABS}. Similarly for $n$ we obtain 
\begin{equation}
\beta(2\alpha+n-n_{\alpha})(2\alpha-n_{\beta}+n_{\beta\alpha})=\alpha(2\beta+n-n_{\beta})(2\beta-n_{\alpha}+n_{\alpha\beta})\ ,\label{spn10}
\end{equation}
which is also the cross-ratio  equation after a simple field redefintion. For $p$ the situation is a little more 
complicated as we have  
$$ s_\alpha^2 = \alpha(\alpha+pp_\alpha)\ ,\qquad  s_\beta^2 = \beta(\beta+pp_\beta) \ ,  $$  
and knowledge of $p_\alpha$ only determines $s_\alpha$ up to a sign. As a result, for given $p,p_\alpha,p_\beta$ there are 
$4$ possibilities for $p_{\alpha\beta}$, which are 
given by solutions of the two multiquadratic quad-graph equations 
\begin{eqnarray}
4\alpha\beta(\alpha-\beta)(p-p_{\alpha\beta})(p_\alpha-p_\beta)   + \alpha\beta(p-p_{\alpha\beta})^2(p_\alpha-p_\beta)^2  
&&   \nonumber  \\ 
+\alpha(\alpha-\beta)(pp_\alpha-p_\beta p_{\alpha\beta})^2+ \beta(\beta-\alpha)(pp_\beta -p_\alpha p_{\alpha\beta})^2 &=& 0 \ ,  \\ 
-4\alpha\beta(\alpha+\beta)(p+p_{\alpha\beta})(p_\alpha+p_\beta)  - \alpha\beta(p+p_{\alpha\beta})^2(p_\alpha+p_\beta)^2  
&&   \nonumber  \\ 
+\alpha(\alpha+\beta)(pp_\alpha-p_\beta p_{\alpha\beta})^2+ \beta(\beta+\alpha)(pp_\beta -p_\alpha p_{\alpha\beta})^2 &=& 0  \ . 
\end{eqnarray}
The first of these is precisely the H3* equation in the Atkinson-Nieszporski classification of integrable 
multiquadratic quad graph equations \cite{AN}, as is the second after a simple field redefinition. 

For $u$ and $v$ we have not succeeded to write a single superposition principle not involving any 
of the other fields. However, using the relations (\ref{sp1})-(\ref{sp4}) it is possible to write the 
following superposition principles involving, respectively, just $u$ and $n$, and just $v$ and $n$: 
\begin{equation} 
\alpha \left( \frac{u_{\alpha}     +u + 2 \alpha}{n_\alpha-n}   - \frac{u_{\alpha\beta}+u_{\beta} + 2\alpha}{n_{\alpha\beta}-n_\beta} \right) 
- \beta \left( \frac {u_{\beta}      +u + 2 \beta}{n_\beta-n}   - \frac{u_{\alpha\beta}+u_{\alpha} + 2\beta} {n_{\alpha\beta}-n_\alpha} \right) = 0 
\end{equation}
\begin{eqnarray} 
&&   \frac{  (v_{\beta}-v)(n_{\beta}-n)      +  (v_{\alpha\beta}-v_{\alpha})(n_{\alpha\beta} - n_\alpha)   }{\beta} 
 - \frac{  (v_{\alpha}-v)(n_{\alpha}-n)    +  (v_{\alpha\beta}-v_{\beta})(n_{\alpha\beta}-n_\beta)}{\alpha} \nonumber\\
&=& 8(\beta- \alpha) \ .
\end{eqnarray}
Here the fields $n$ satisfy the cross-ratio type equation (\ref{spn10}). 

\section{Travelling wave solutions}
In this section we apply the BT (\ref{BT4}),(\ref{BT3}) where $s$ satisfies (\ref{BT1}),(\ref{BT2})
to the constant solution of CH $u=u_0\not=0$, to obtain travelling wave solutions, specifically soliton 
and cuspon solutions. These and other travelling wave solutions have been extensively studied in the 
literature, see for example \cite{LO1,LO2,Boyd1,Boyd2,KZ1,Park1,Park2,PARKER1,Len1,PV0}, and  
the BT is just one of many methods to derive them. The advantages of the BT will become apparent when we study 
superposition in the next section. 

If $\alpha(\alpha+u_0)>0$ 
there are two kinds of real solutions of (\ref{BT1}),(\ref{BT2}): 
\begin{equation}
s_{\alpha}= \sqrt {\alpha \left(u_0  +\alpha\right) }
\tanh \left( \frac{\sqrt{\alpha \left(u_0+\alpha \right) }\left(x -  x_0 + \left(2\alpha-u_0 \right) t \right) }{2\alpha} \right)\ ,
\label{s1}\end{equation}
which we call the ``tanh-type'' solution, and the same with tanh replaced by coth, which we call the ``coth-type'' solution. 
As we will see both of these give rise to travelling wave solutions. If $\alpha(\alpha+u_0)<0$ then there are real solutions 
\begin{equation}
s_{\alpha}= \sqrt {-\alpha \left(u_0  +\alpha\right) }
\tan \left( \frac{\sqrt{-\alpha \left(u_0+\alpha \right) }\left(x -  x_0 + \left(2\alpha-u_0 \right) t \right) }{2\alpha} \right)\ ,
\end{equation}
and the same with tan replaced by cot, and an overall minus sign. Both of these
give rise to periodic solutions (see for example \cite{Boyd1,Len1}), but these will not be studied here. 

Returning to the case $\alpha(\alpha+u_0)>0$, it is useful to
write $\alpha+u_0=\alpha  U^2$, where $U>0$, so the solution 
(\ref{s1}) becomes 
\begin{equation}
s_{\alpha}= \alpha U
\tanh \left( \frac{U}{2} \left(x -  x_0 + \left(3-U^2 \right)\alpha  t \right)   \right)\ ,
\label{s1n}\end{equation}
and the same with coth for a coth-type solution. 
Using  (\ref{BT4}),(\ref{BT3}) the resulting solution is $u_\alpha(x_\alpha,t)$ where 
\begin{eqnarray}
u_\alpha &=&     
\alpha( U^2-3) +   \frac {2 \alpha(U^2-1) }{U^2  \tanh^2 \left( \frac{U}{2} \left(x -  x_0 + \left(3-U^2 \right)\alpha  t  \right)\right)  -1 } 
 \ ,\label{UU1}\\
x_\alpha &=&   x + \ln \left| \frac {U\tanh \left( \frac{U}{2}\left(x -  x_0 + (3-U^2)\alpha  t  \right)\right) -1  }
                                   {U\tanh \left( \frac{U}{2}\left(x -  x_0 +(3-U^2)\alpha  t   \right)\right) +1  } \right|  \label{UU2}
\end{eqnarray}
or the same with coth. Finally, 
writing $z=x-x_0 + \left(3-U^2 \right)\alpha  t$,  the solution becomes  $u_\alpha(x_\alpha,t)$ where 
\begin{eqnarray}
u_\alpha &=&   
\alpha( U^2-3) +   \frac {2 \alpha(U^2-1) }{U^2  \tanh^2 \frac12 U z   -1 } 
 \ ,\label{UU1n}\\
x_\alpha-x_0 + (3-U^2)\alpha t  &=&   z + \ln \left| \frac {U\tanh \frac12 Uz -1  }{U\tanh \frac12 Uz +1  } \right|\ ,   \label{UU2n}
\end{eqnarray}
this being a tanh-type solution, or a coth-type solution, which is the same with tanh replaced by coth. 
Both tanh-type and coth-type solutions are 
travelling waves with speed $c=(U^2-3)\alpha$, written in an implicit form. The first step in analyzing these
solutions is to decide whether the maps from $x_\alpha$ to $z$ are bijections. For {\underline{ tanh-type 
solutions with $U<1$}},  neither the factor in the numerator 
or in the denominator inside the $\ln$ can vanish, and thus $x_\alpha$ only tends to (plus or minus) infinity as $z$ tends to 
(plus or minus) infinity. The corresponding solutions are solitons which tend to $u_0=\alpha(U^2-1)$ at spatial
infinity, with speed $c=\alpha(U^2-3)$ and central elevation $-\alpha(1+U^2)=c-2u_0$. Note that since 
$$   U = \sqrt{\frac{3u_0-c}{u_0-c}} $$ 
and $0<U<1$ we must either have $c<3u_0<0$  or $0<3u_0<c$.  
Figure 1 displays the soliton profile for $c=2$ and $u_0=0.5,0.1,0.02$. 
(For negative $u_0$ and $c$ the soliton is  inverted.) 

\begin{figure}[ht]
\begin{center}
\includegraphics[height=3in]{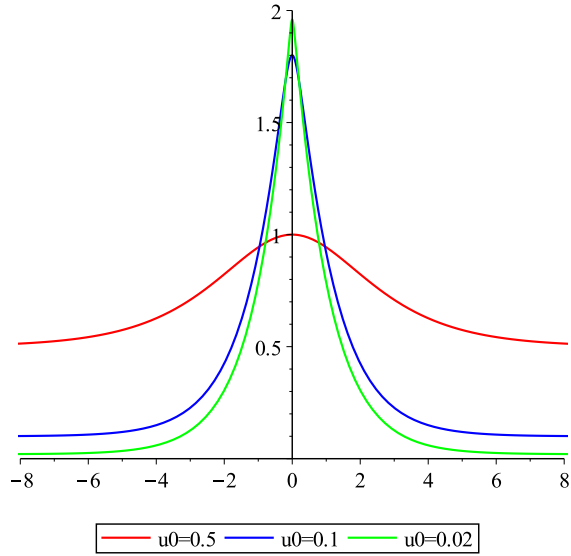}
\end{center}
\caption{Soliton profile, $c=2$, $u_0=0.5,0.1,0.02$.} 
\end{figure}

Of particular interest is the limit of the soliton for fixed $c$ and $u_0\downarrow 0$ (for $c>0$) or
$u_0\uparrow 0$ (for $c<0$). Figure 2 shows $x_\alpha$ as a function of $z$ (for $x_0=t=0$), $u_\alpha$ as a function of $z$ and 
$u_\alpha$ as a function of $x_\alpha$ in the case $c=2$, $u_0=10^{-8}$. $x_\alpha$ is close to zero, and $u_\alpha$ is close to $c$
for a large interval of $z$ values of size $O\left(\left|\ln(u_0/c)\right|\right)$ around $z=0$. In the plot of $u_\alpha$ against $x_\alpha$
this gives rise to a sharp peak. This is the peakon limit. To see this analytically it is possible to use  
(\ref{UU1n}) to find $z$ in terms of $u$ (with a $\pm$ uncertainty as it is necessary to take a square root), 
and then (\ref{UU2n}) becomes
\begin{eqnarray}
x_\alpha-x_0 + (3-U^2)\alpha t & =& \pm \left( 
2 \sqrt{\frac{c-u_0}{c-3u_0}} {\rm arctanh} \left( \sqrt{\frac{c-u_0}{c-3u_0}}  \sqrt{1-\frac{2u_0}{c-u_\alpha}}  \right) 
\right. \nonumber \\
&& +\left.  \ln\left( \frac{c-u_\alpha}{2u_0}  \left( 1 -  \sqrt{1-\frac{2u_0}{c-u_\alpha}}  \right)^2   \right) 
   \right) \ . 
\end{eqnarray}
Both terms on the RHS diverge as $u_0\rightarrow 0$, but it is straightforward to extract the divergent behavior, which 
cancels between the terms, and to obtain the limit, which is simply $\pm \ln\left(\frac{u_\alpha}{c}\right)$. 

\begin{figure}[ht]
\begin{center}
\includegraphics[height=2in]{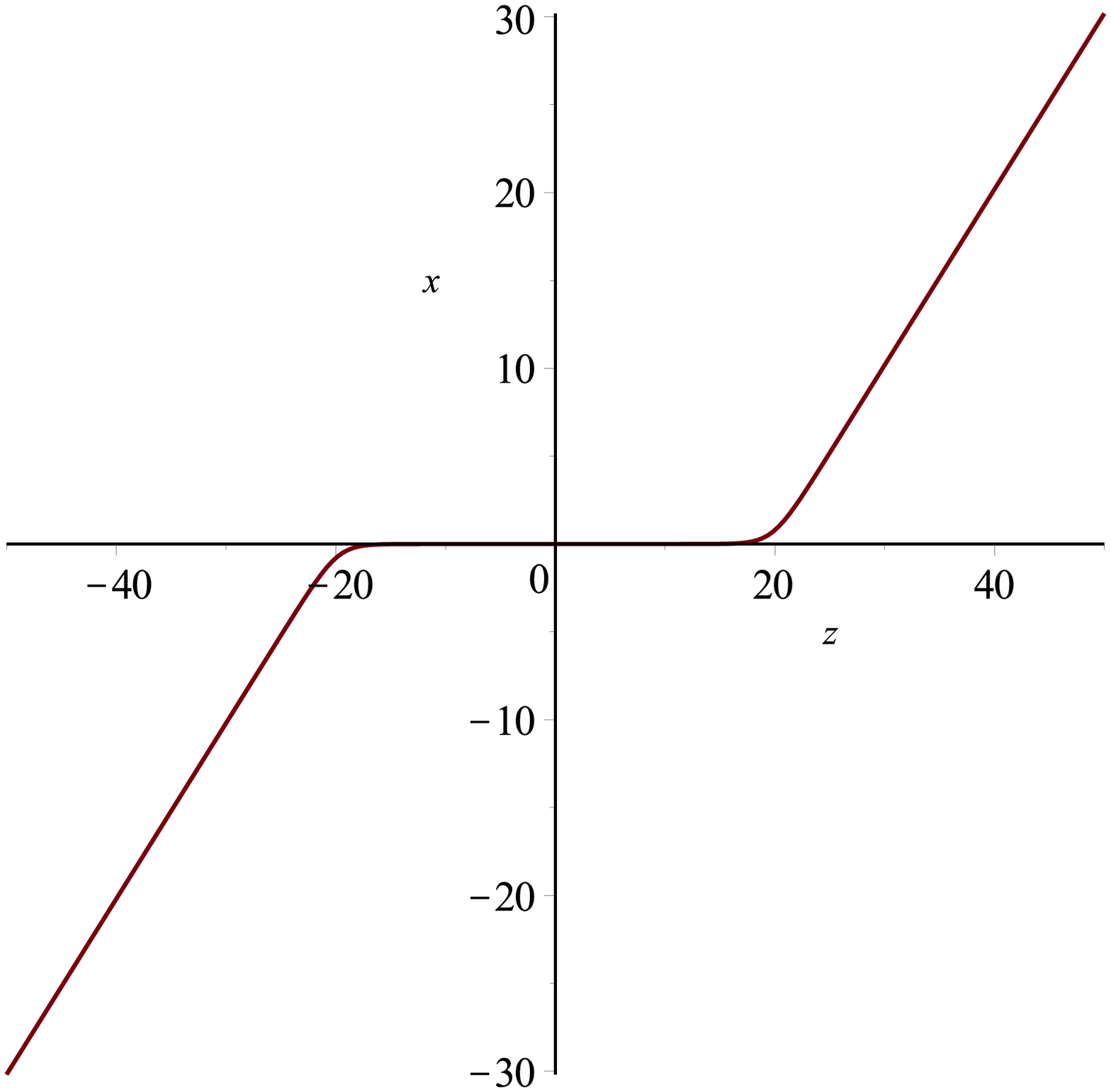}
\includegraphics[height=2in]{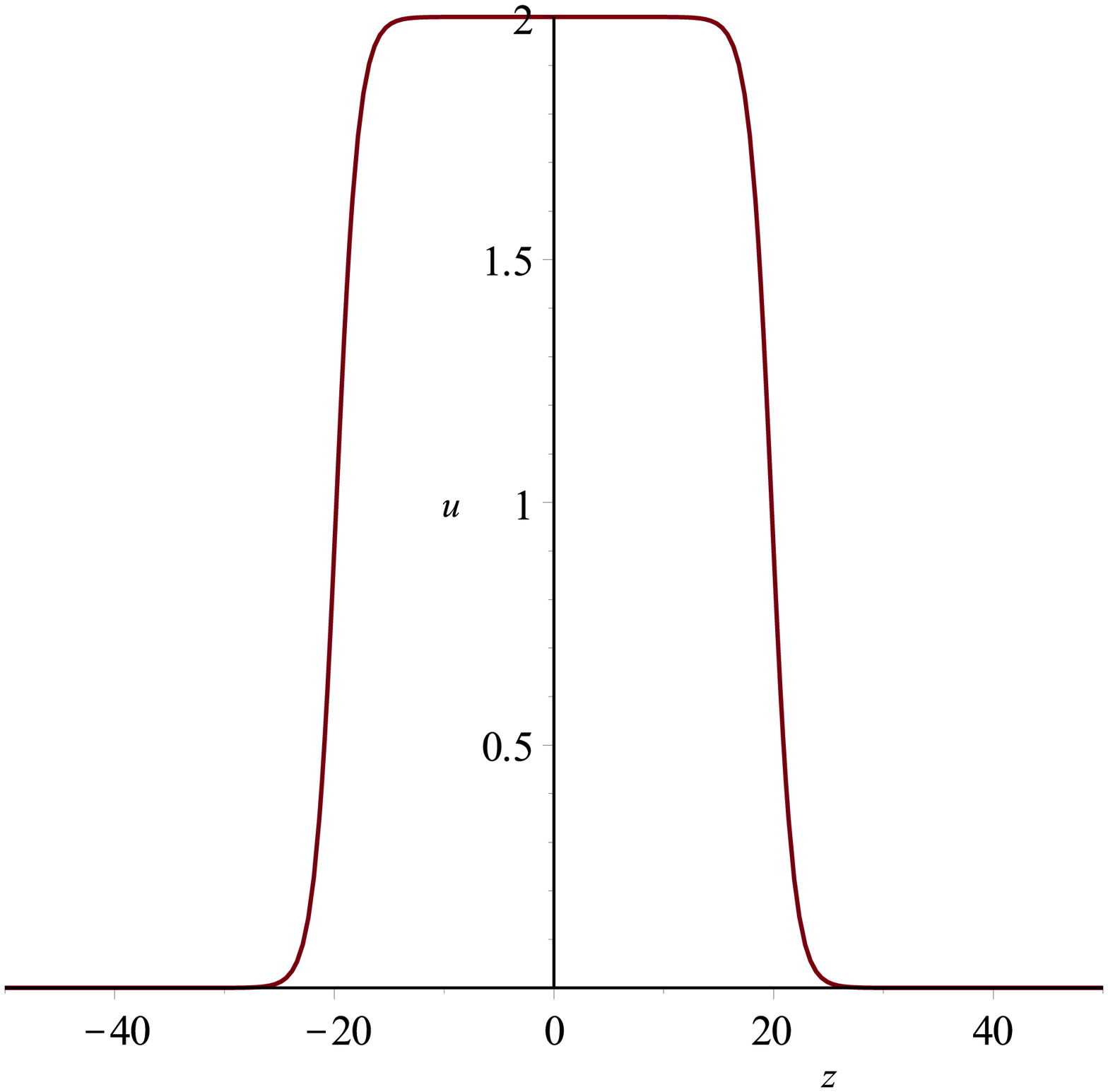}
\includegraphics[height=2in]{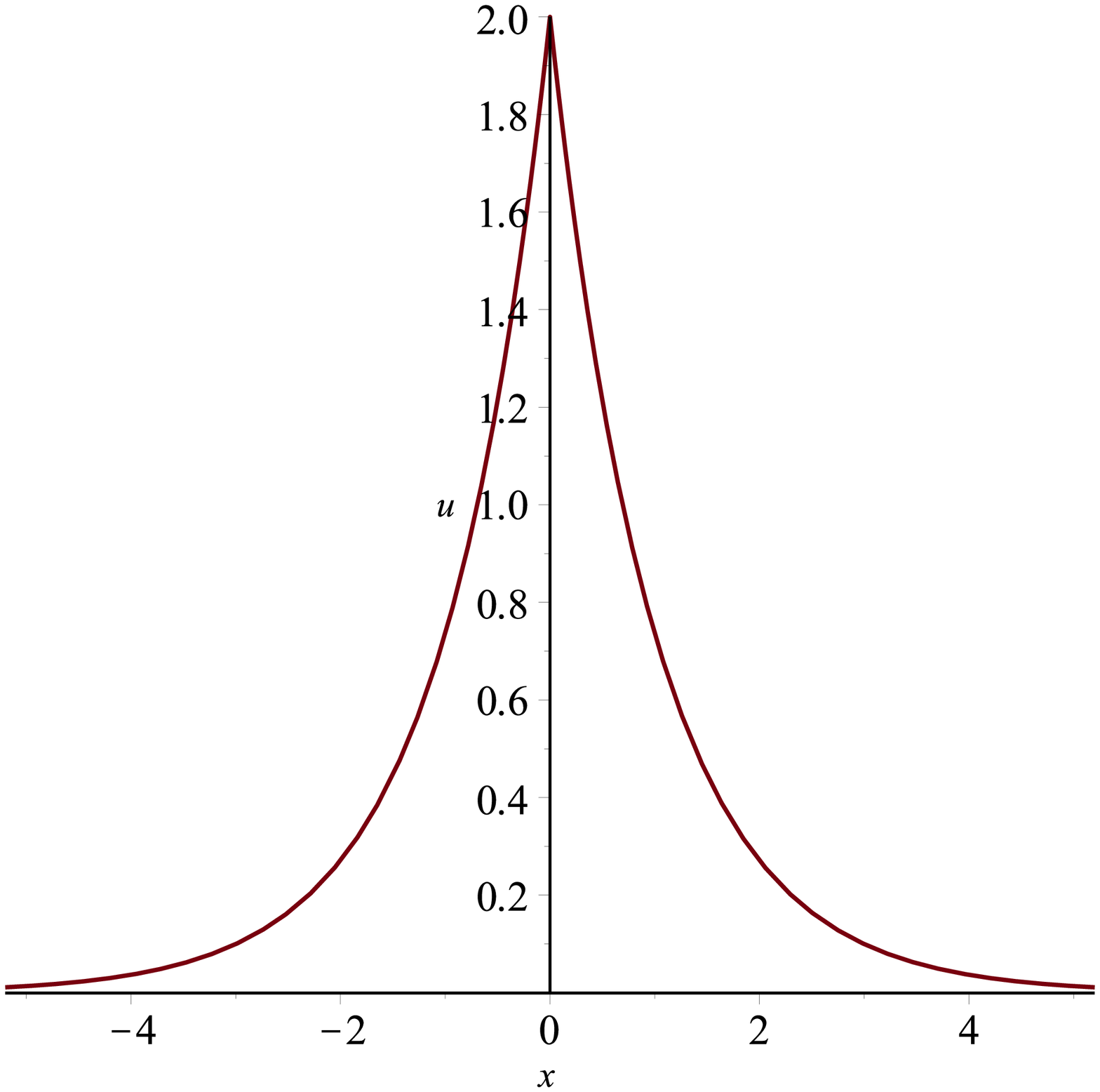}
\end{center}
\caption{The soliton with $c=2$, approaching the peakon limit. 
$u_0=10^{-8}$. $x$ as a function of $z$, $u$ as a function of $z$ and $u$ as a function of $x$.} 
\end{figure}

Moving now to {\underline{ tanh-type  solutions with $U>1$}},  from (\ref{UU2n}) we expect $x_\alpha$ to diverge when 
$\tanh \frac12 U z = \pm \frac1{U} $ and thus the map from $z$ to $x_\alpha$ will not be a bijection. 
Figure 3 shows $x_\alpha$ and $u_\alpha$ as functions of $z$ for $c=2$ and $u_0=3$. The map from $z$ to $x_\alpha$ is
$3$ to $1$ and thus there are $3$ corresponding solutions of CH, depicted in Figure 4. Since these are all unbounded we
do not devote further attention to them. 

\begin{figure}[ht]
\begin{center}
\includegraphics[height=3in]{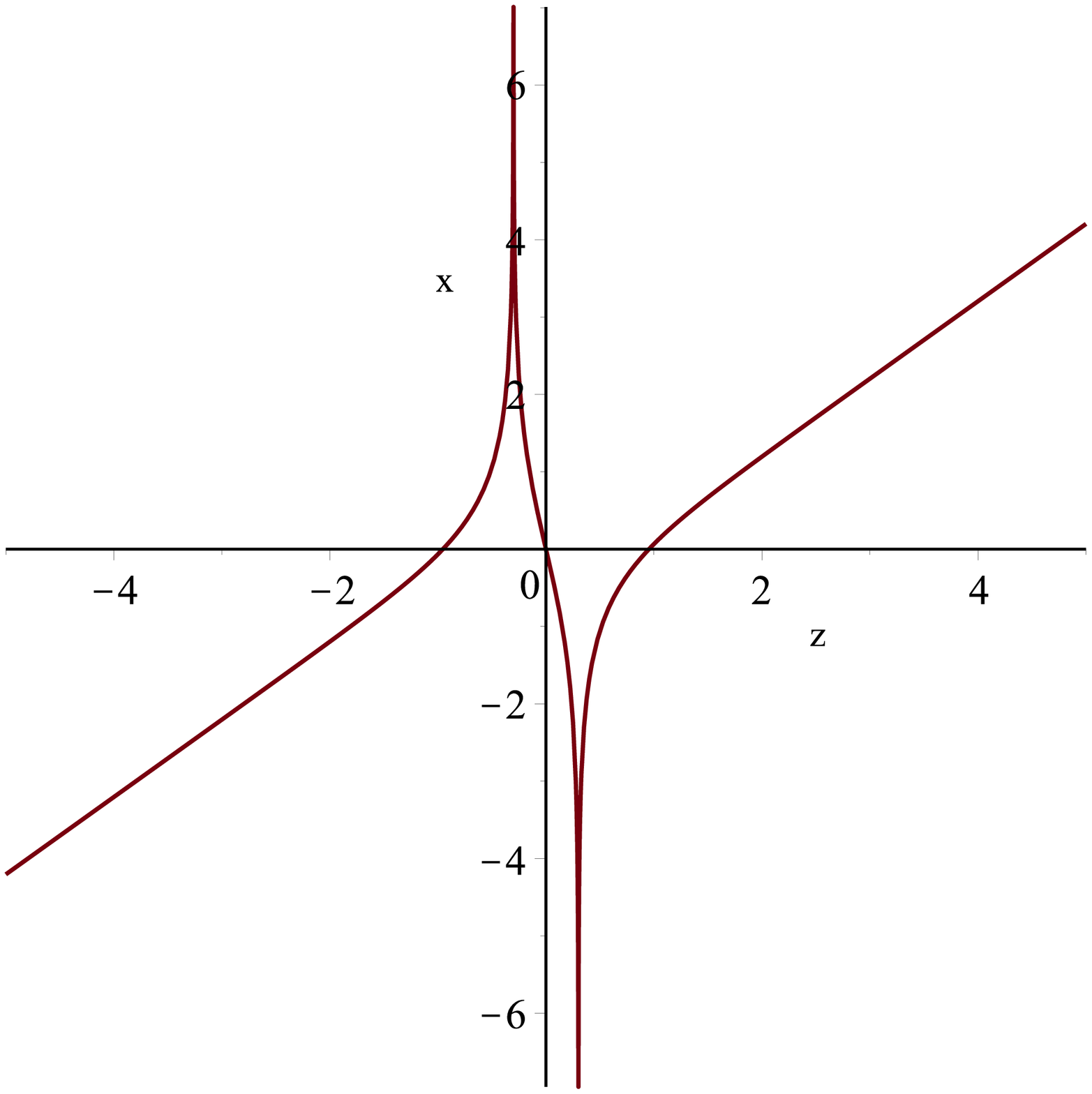}
\includegraphics[height=3in]{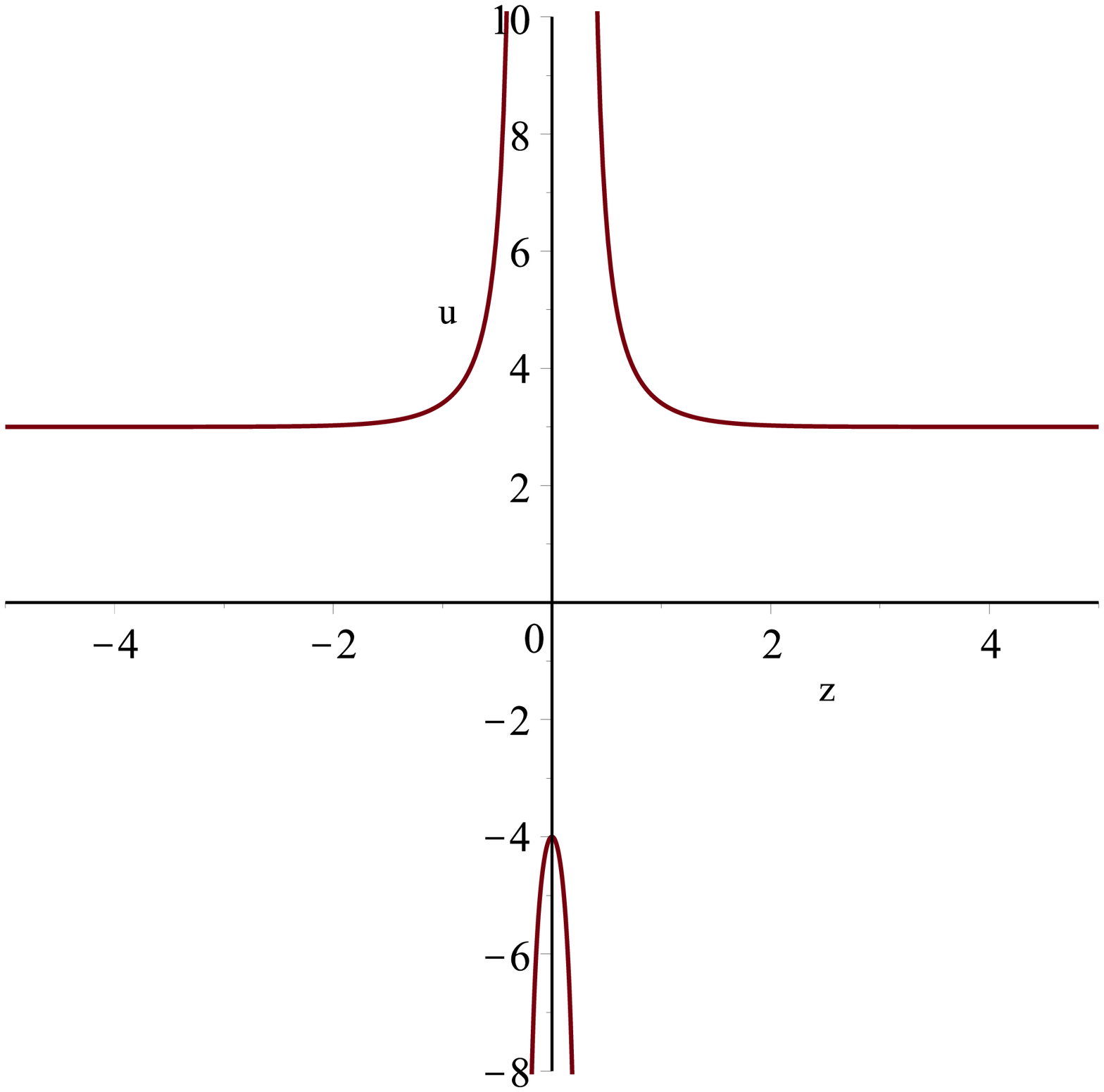}
\end{center}
\caption{tanh-type solutions with $c=2$ and $u_0=3$ ($U=\sqrt{7}>1$), $x_\alpha$ and $u_\alpha$ as functions of $z$.
The map from $z$ to $x_\alpha$ is not $1-1$.}
\end{figure}

\begin{figure}[ht]
\begin{center}
\includegraphics[height=2in]{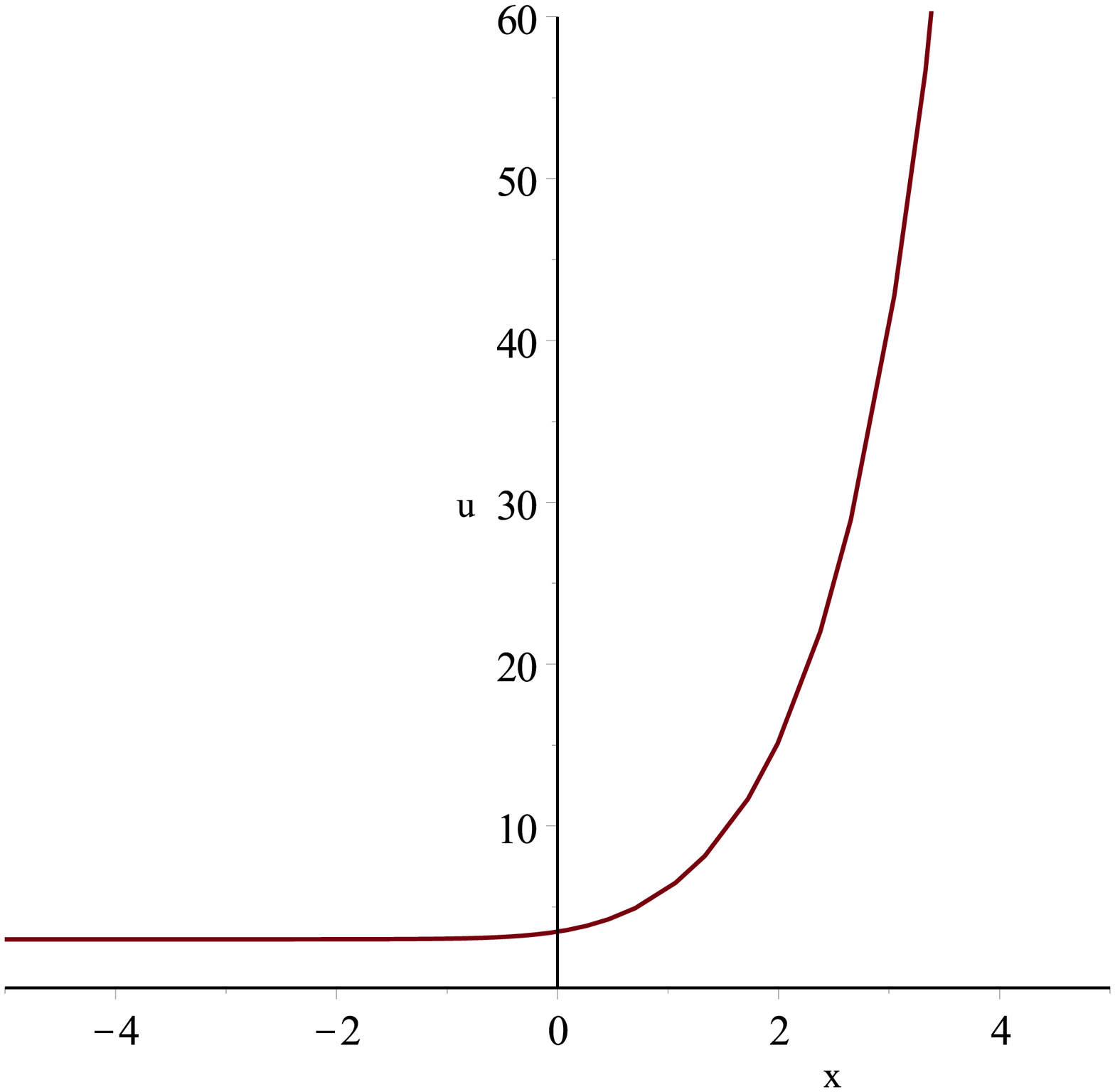}
\includegraphics[height=2in]{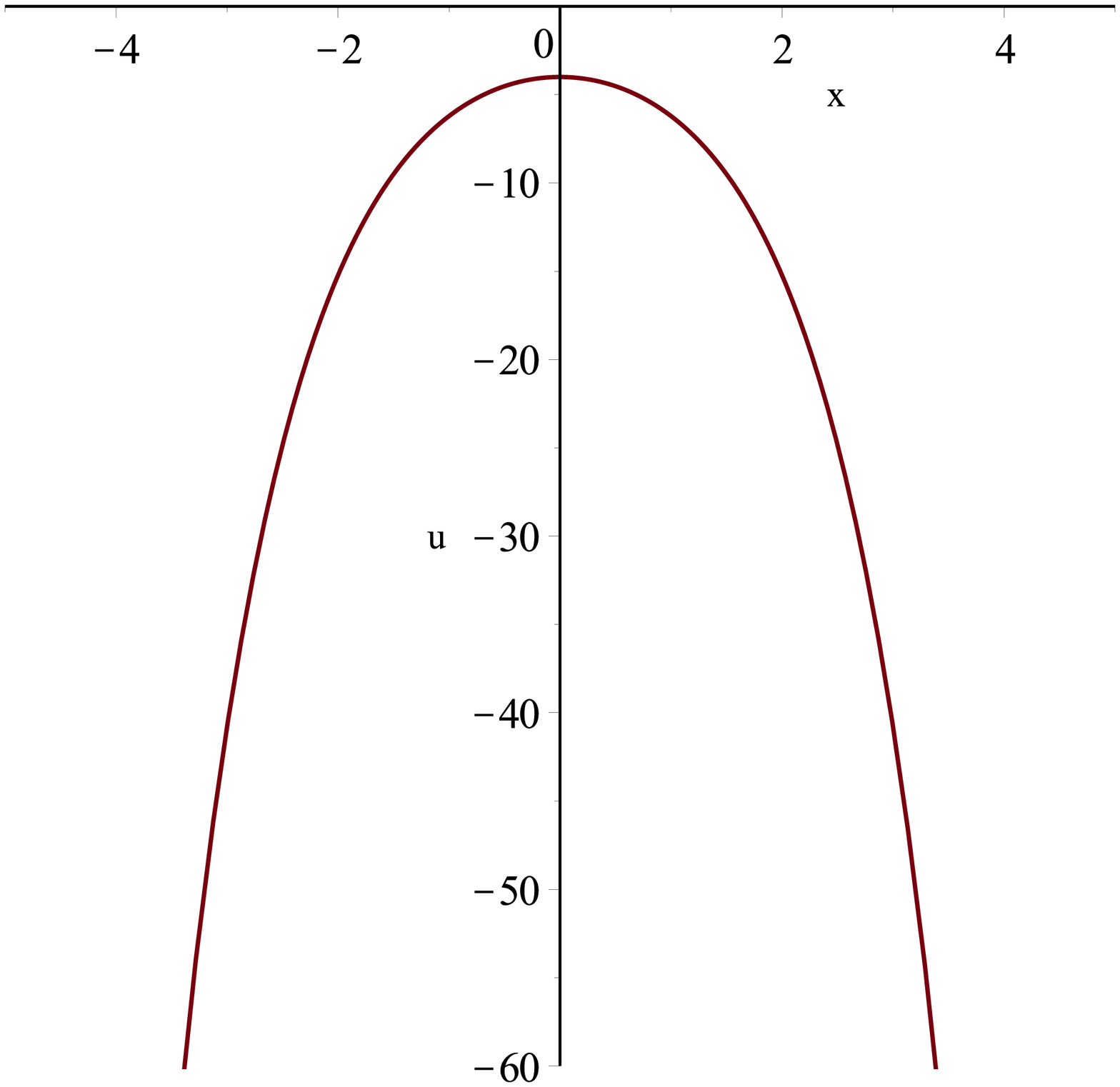}
\includegraphics[height=2in]{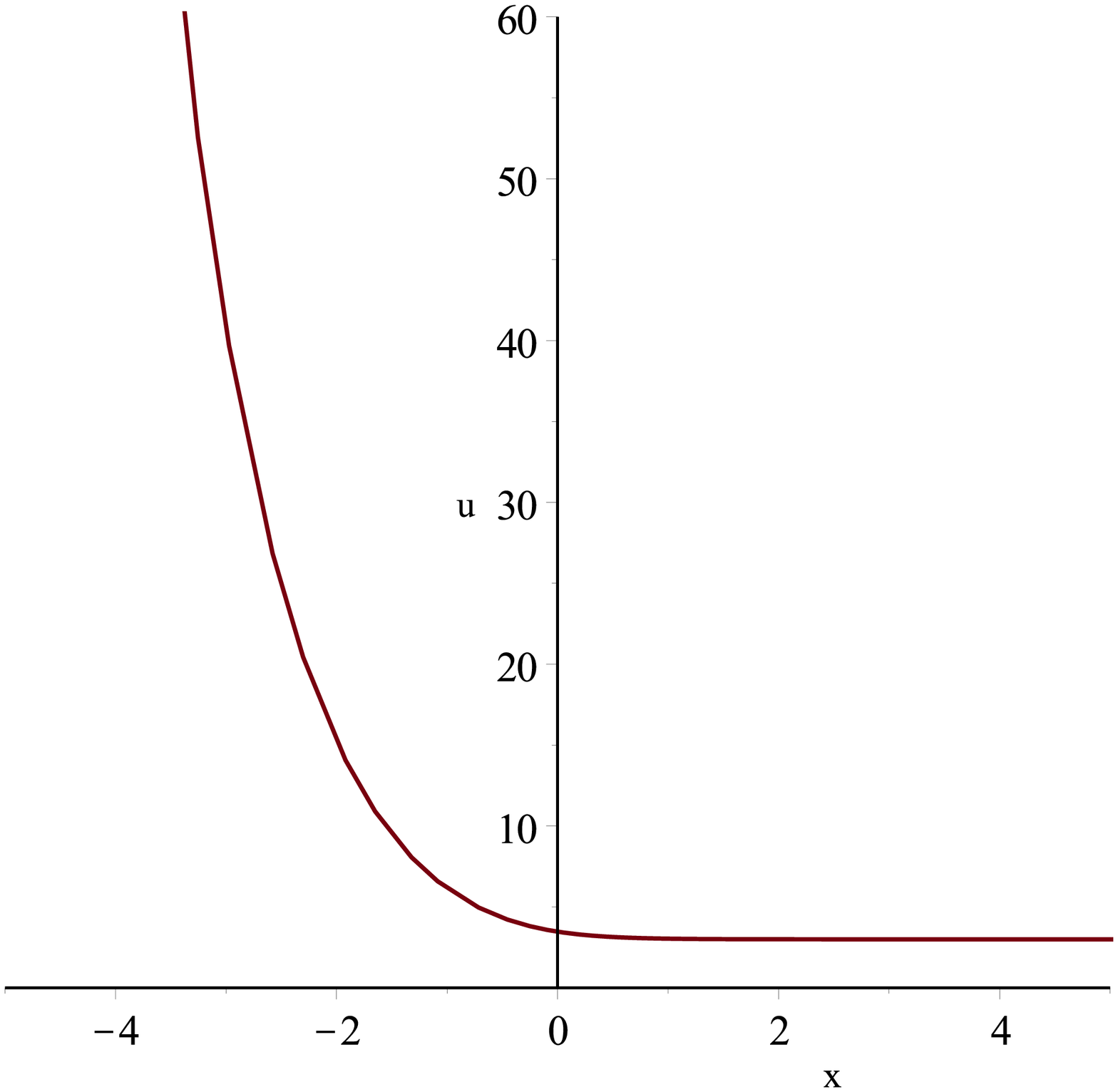}
\caption{tanh-type solutions with $c=2$ and $u_o=3$ ($U=\sqrt{7}>1$), $u_\alpha$ as a function of $x_\alpha$ for 
the $3$ unbounded solutions of CH.}
\end{center}
\end{figure}

Moving now to coth-type solutions, the situation is very similar, but now the map from $z$ to $x_\alpha$ 
will be $1-1$ if $U>1$ and many to $1$ if $U<1$, and there is  a subtlety arising due to the divergence of 
$\coth\frac12 Uz$ at $z=0$. 
For  {\underline{ coth-type  solutions with $U<1$}},  $x_\alpha$  diverges when 
$\coth \frac12 U z = \pm \frac1{U} $. The map from $z$ to $x_\alpha$ is once again $3$ to $1$.  
Figure 5 shows $x_\alpha$ and $u_\alpha$ as functions of $z$ for $c=2$ and $u_0=\frac12$, and Figure 6 shows 
the $3$ corresponding solutions of CH. The subtlety, as can be seen in Figure 7,  is that the solution 
corresponding to the range of $z$'s that includes zero, has a cusp at $z=0$, arising from the divergence of 
$\coth\frac12 Uz$. Since at this point ${u}_\alpha$ is not differentiable, it is necessary to ask in what
sense this is a solution of CH. Fortunately, the value of ${u}_\alpha$ at the cusp 
is $c$, which makes it possible to interpret
the solution in a weak sense \cite{Len1}, though we do not go into details here. 

\begin{figure}[ht]
\begin{center}
\includegraphics[height=3in]{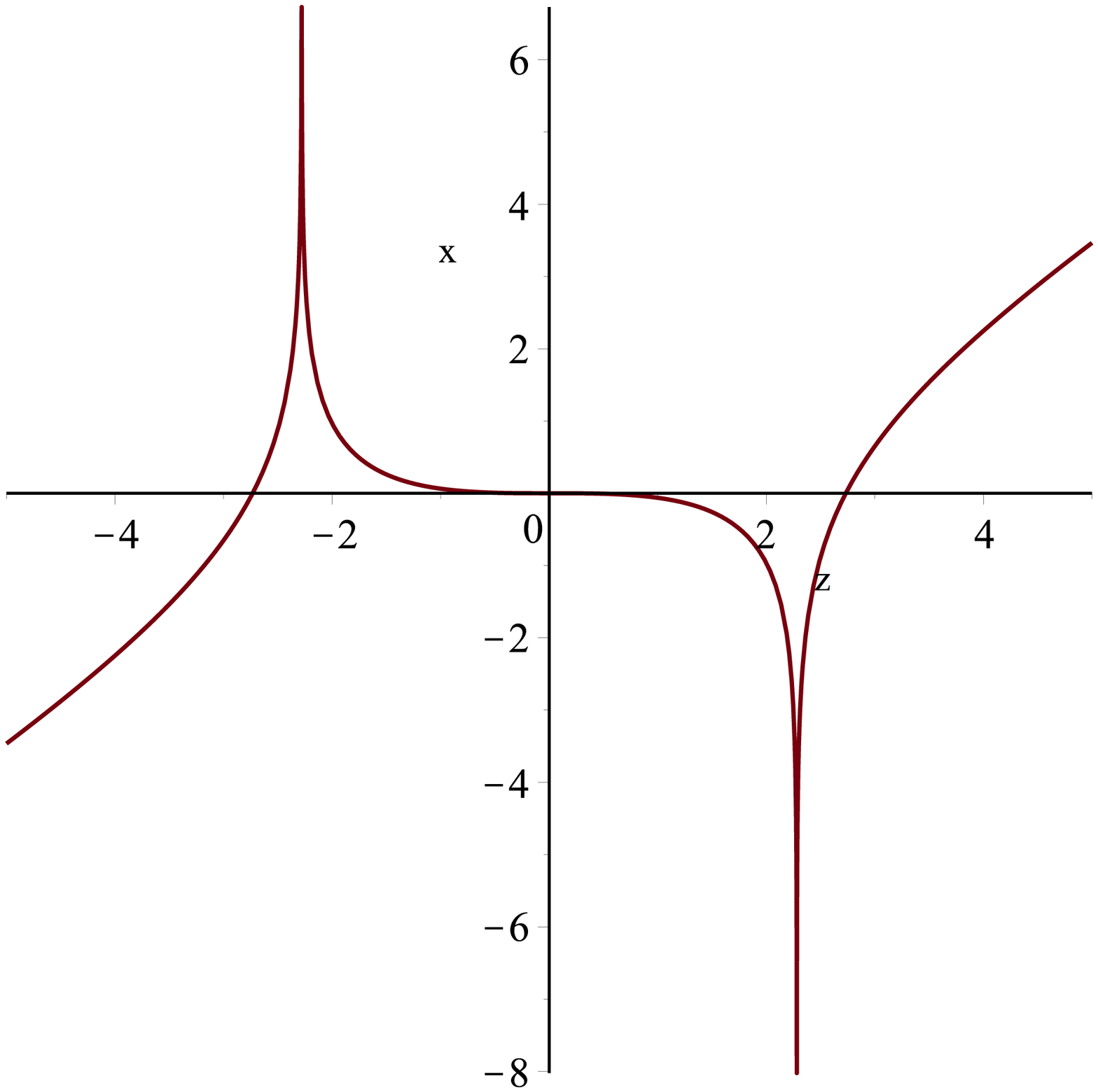}
\includegraphics[height=3in]{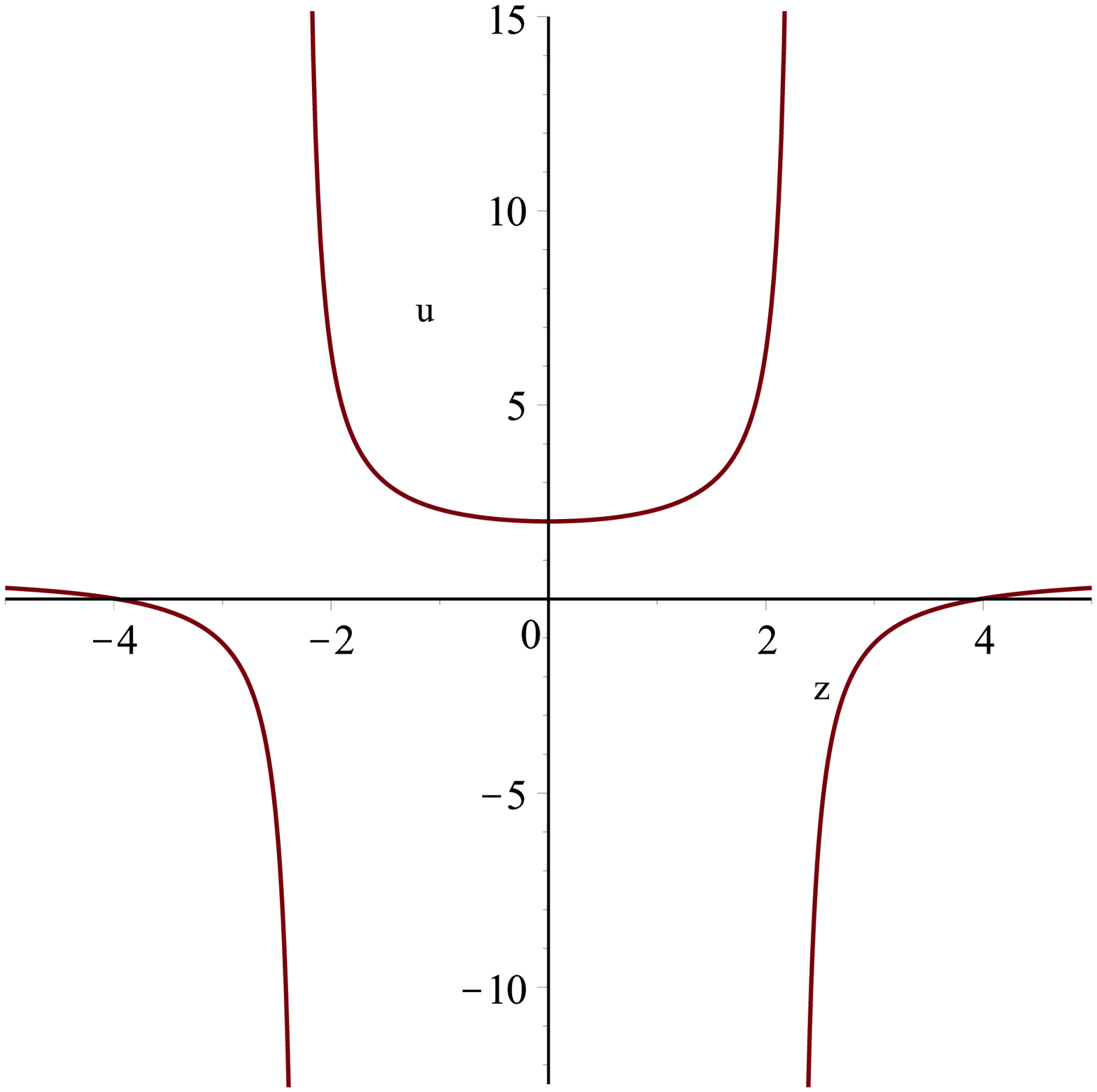}
\end{center}
\caption{coth-type solutions with $c=2$ and $u_0=\frac12$ ($U=\frac1{\sqrt{3}}<1$), 
${x}_\alpha$ and ${u}_\alpha$ as functions of $z$.
The map from $z$ to ${x}_\alpha$ is not $1-1$.}
\end{figure}

\begin{figure}[ht]
\begin{center}
\includegraphics[height=2in]{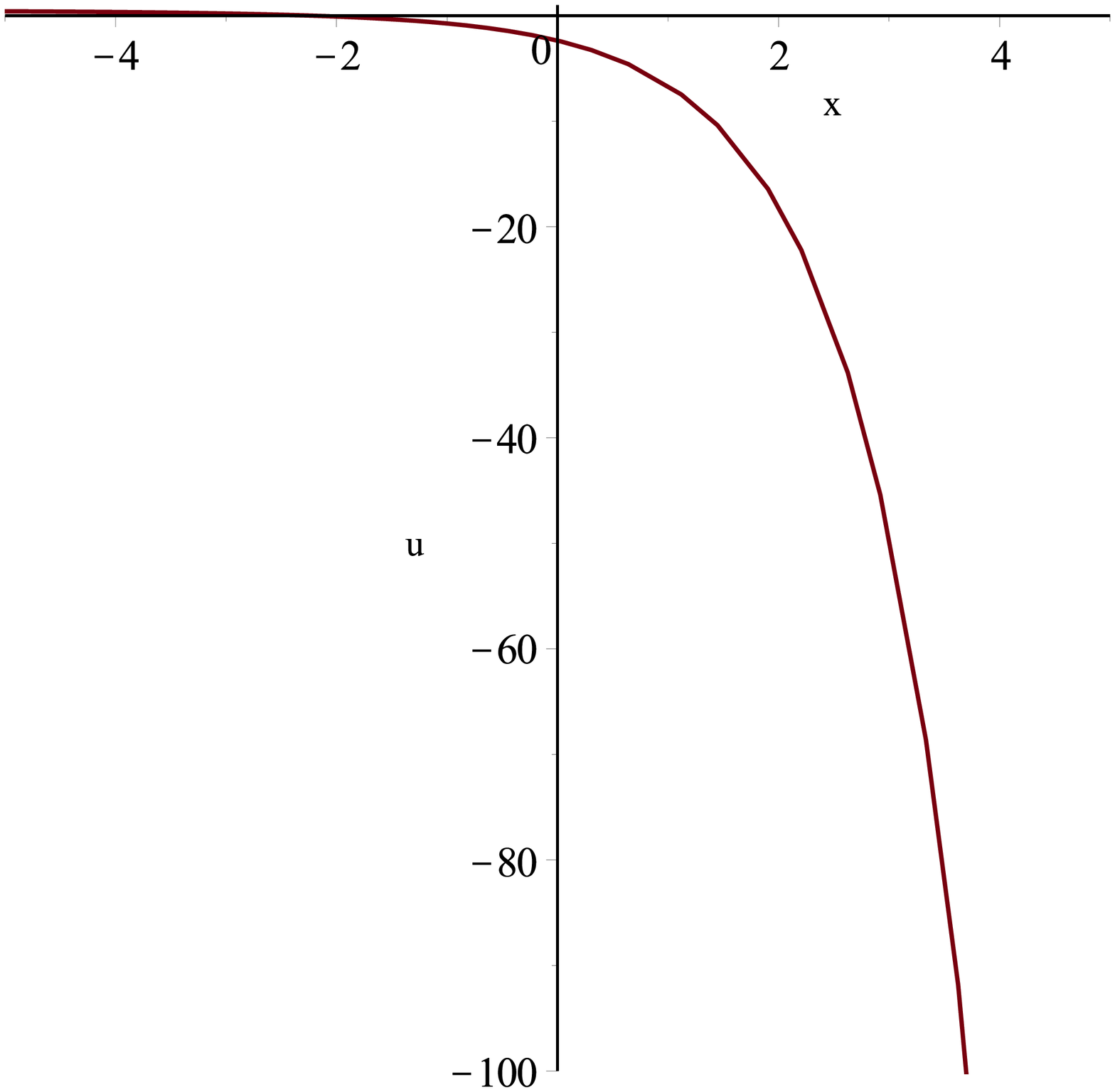}
\includegraphics[height=2in]{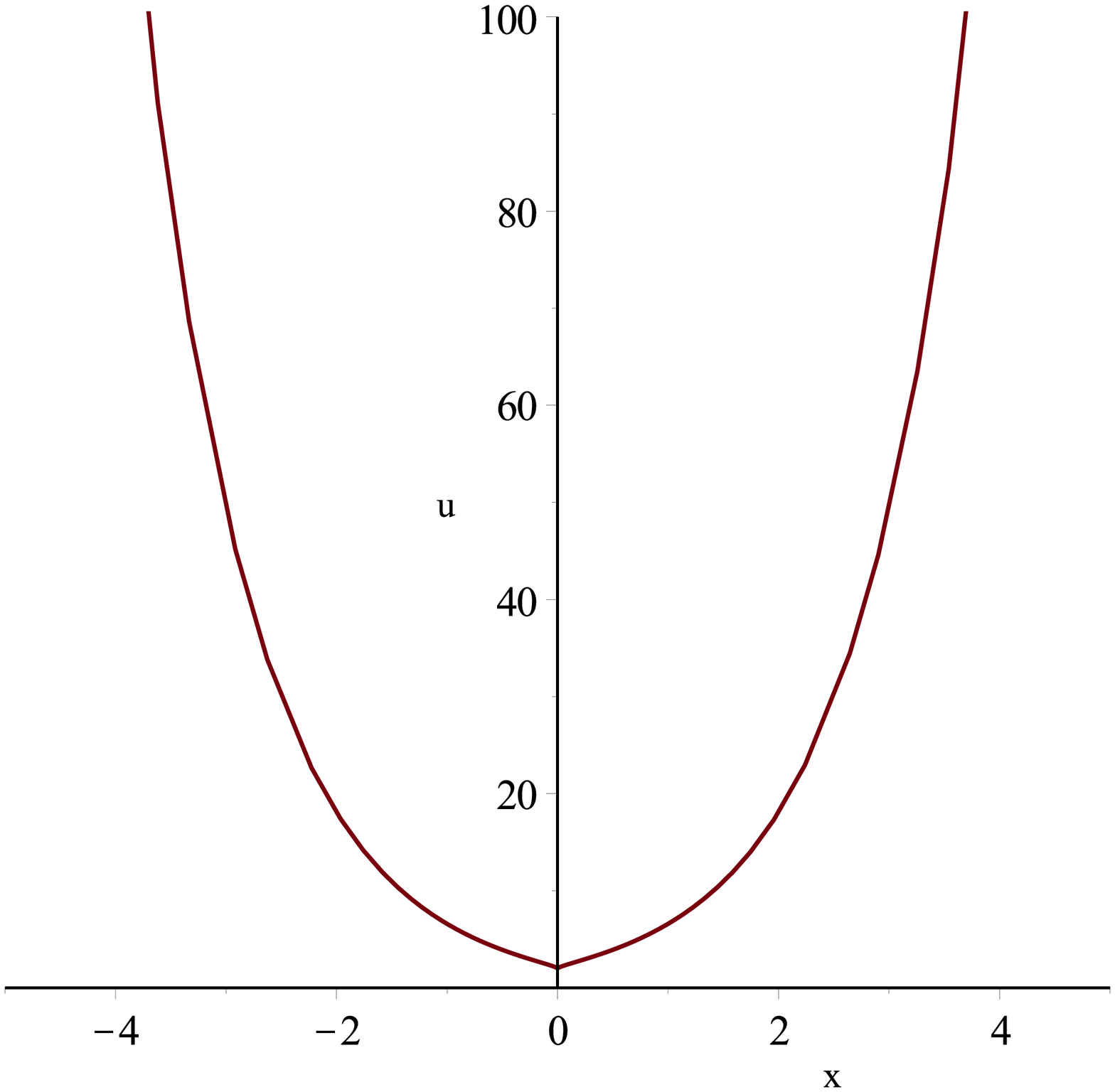}
\includegraphics[height=2in]{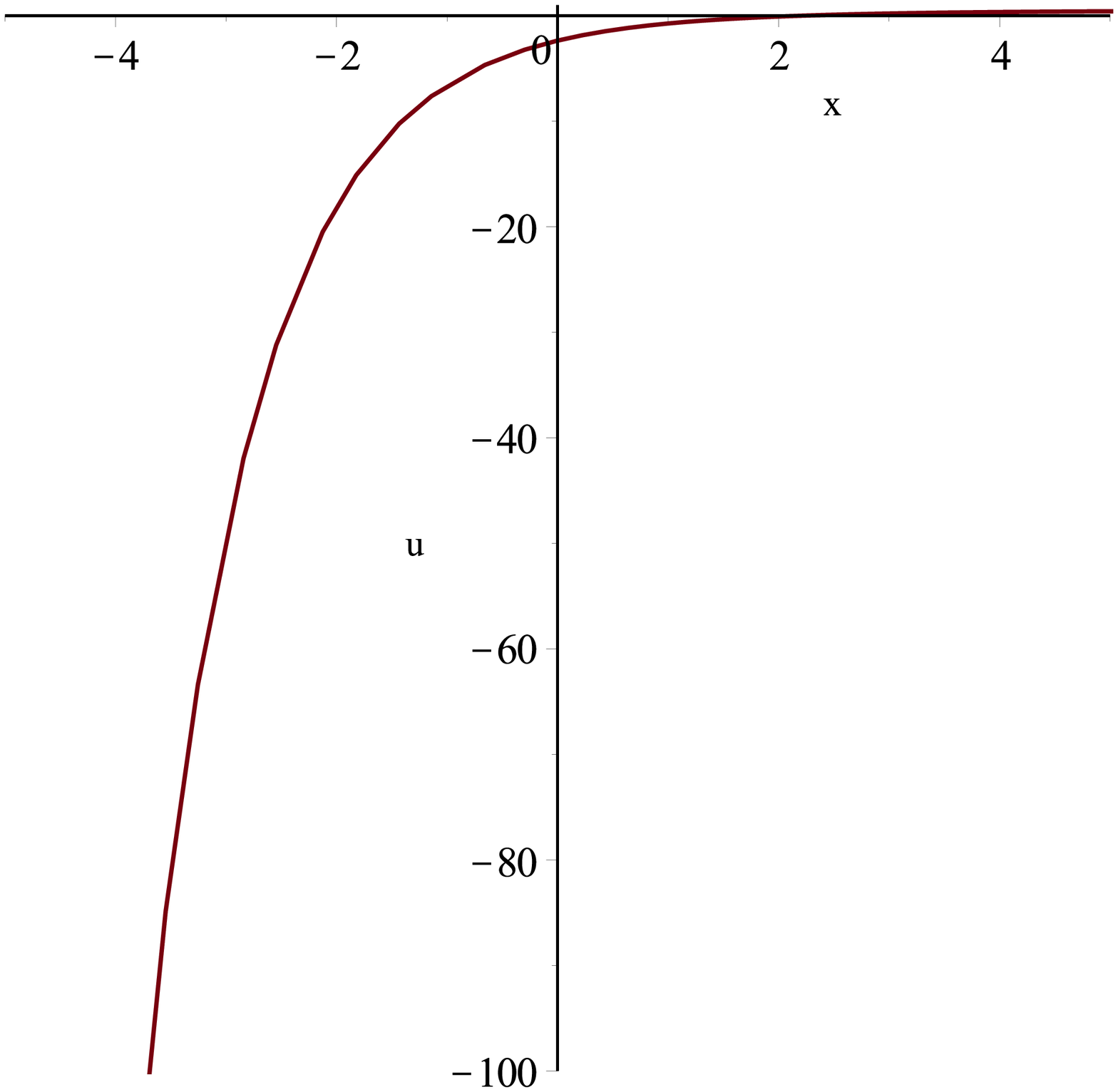}
\caption{coth-type solutions with $c=2$ and $u_0=\frac12$ ($U=\frac1{\sqrt{3}}<1$), 
${u}_\alpha$ as a function of ${x}_\alpha$ for 
the $3$ unbounded solutions of CH.}
\end{center}
\end{figure}

\begin{figure}[ht]
\begin{center}
\includegraphics[height=2in]{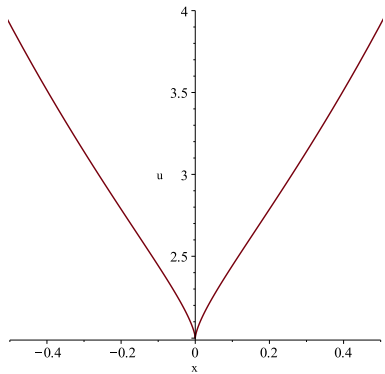}
\caption{coth-type solutions with $c=2$ and $u_0=\frac12$ ($U=\frac1{\sqrt{3}}<1$), 
close up on the cusp in one of the solutions of CH.}
\end{center}
\end{figure}

For  {\underline{ coth-type  solutions with $U>1$}},  the map from $z$ to ${x}_\alpha$ is a bijection, and 
once again there is a single solution of CH, but with a cusp at $z=0$ --- this is the cuspon solution. Due to 
the requirement $U>1$ cuspon solutions only exist with speed  $c<u_0$ if $u_0$ is positive, and speed $c>u_0$ if
$u_0$ is negative. Figure 8 illustrates cuspon solutions with $c=2$ for $u_0=-1,-0.5,-0.1$. 
(For positive $u_0$  the cuspon is  inverted.) 
Note that the central elevation of the cusp is $c$, as required for it to be a weak solution. For $c>0$ ($c<0$)
it is possible to consider the limit of the cuspon as $u_0\uparrow 0$ ($u_0\downarrow 0$), and this is 
once again the peakon limit. 

\begin{figure}[ht]
\begin{center}
\includegraphics[height=3in]{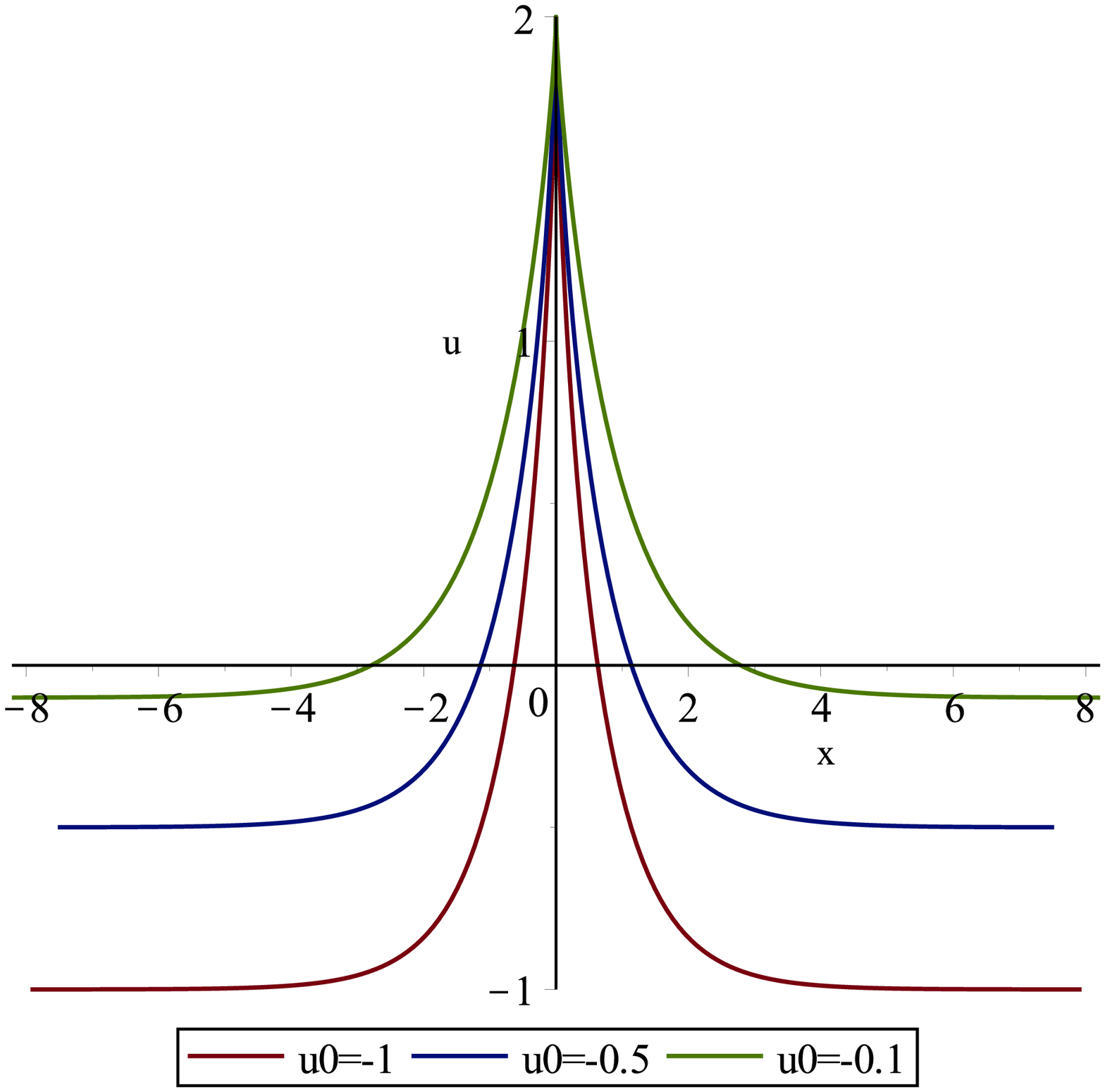}
\end{center}
\caption{Cuspon profile, $c=2$, $u_0=-1,-0.5,-0.1$.} 
\end{figure}

\newpage 
We summarize the travelling waves presented in this section in the following table. All the solutions  have 
asymptotic height $u_0$:

\vskip.3in 

\begin{center}
\begin{tabular}{ l | l | l | l }
\hline
tanh-type   &   $U<1$   &  soliton   &   
                 central elevation $c-2u_0$   \\ 
 &&&                $u_0>0$:  $c>3u_0$        \\
 &&&                 $u_0<0$:  $c<3u_0$  inverted    \\
                                    &    $U>1$   &  unphysical &      \\   \hline
coth-type   &  $U<1$ &  unphysical  &      \\
                                      &  $U<1$  & cuspon   &  
                central elevation $c$    \\
&&&                $u_0>0$:  $c<u_0$   inverted  \\
 &&&               $u_0<0$:  $c>u_0$    \\
\hline 
\end{tabular}
\end{center}

\newpage
\section{Two wave solutions}

The first investigations of two wave solutions were \cite{Je1} and \cite{KZ2}, both of which required some element
of numerical computation.  However, since then, a substantial literature 
\cite{Johnson,LZ1,Park3,Dai1,Li,MAT2,PM1,MAT1,PARKER3,Dai3,XZQ} has developed 
on  multisoliton, multicuspon and soliton-cuspon solutions. 
The known methods for analytic construction of solutions include a determinantal formula based on the 
inverse scattering approach, a Hirota bilinear form for CH and a reciprocal transformation relating the 
CH hierarchy to the KdV hierarchy.  (For multipeakon solutions very different
techniques are involved \cite{CH0,BSS1,BSS2,BSS4,PARKER2}.) As we will shortly see, use of the superposition principle gives 
a further very simple method. 

In our approach, 
two wave solutions should be 
obtained using formulas (\ref{spu}) and (\ref{spx}), taking $u=u_0$ to be constant and $s_\alpha$ ($s_\beta$) either 
of tanh-type, as given in (\ref{s1n}) where $U=U_\alpha=\sqrt{1+\frac{u_0}{\alpha}}$
($U=U_\beta=\sqrt{1+\frac{u_0}{\beta}}$) and $x_0=x_{\alpha,0}$ ($x_0=x_{\beta,0}$), 
or of coth-type, which is identical  but with coth.  The only question is which superpositions of this type give
maps from $x$ to $x_{\alpha\beta}$ that are 1-1. 

%

\smallskip 

{\noindent \bf Proposition.} The following 3 superpositions give maps from $x$ to  $x_{\alpha\beta}$ which are 1-1: 
\begin{enumerate}
\item tanh-type solutions $s_\alpha$ with $U_\alpha<1$  with tanh-type solutions $s_\beta$ with $U_\beta>1$ (so $\frac{u_0}{ \alpha}<0
<\frac{u_0 }{\beta}$) ---   soliton-cuspon  superpositions. 
\item tanh-type solutions $s_\alpha$ with $U_\alpha<1$  with coth-type solutions $s_\beta$ with $U_\beta<1$, with $U_\alpha<U_\beta$ ---
soliton-soliton  superpositions. 
\item tanh-type solutions $s_\alpha$ with $U_\alpha>1$  with coth-type solutions $s_\beta$ with $U_\beta>1$, with $U_\beta<U_\alpha$ ---
cuspon-cuspon  superpositions. 
\end{enumerate}

\smallskip
 
\noindent
Note here, for example, that a soliton-soliton superposition is {\em not} as we might expect, the superposition of 
two tanh-type solutions with $U<1$, but the superposition of a tanh-type solution with $U<1$ with a unphysical 
coth-type solution with $U<1$. 

\smallskip 

{\noindent \bf Proof.} It is necessary to show in each case 
that neither the numerator or denominator of the expression inside the $\ln$ in (\ref{spx}) vanishes, i.e. 
that $|s_\beta-s_\alpha| \not= |\beta-\alpha|$.  
In the calculations below we repeatedly use the identities
$$ \alpha = \frac{u_0}{U_\alpha^2-1}  \ , \qquad \beta = \frac{u_0}{U_\beta^2-1} .   $$ 
\begin{enumerate}
\item  In this case we have  
\begin{eqnarray*} 
|s_\beta - s_\alpha | 
&=&  \left|  \beta U_\beta \tanh \left( \ldots \right)    -  \alpha U_\alpha  \tanh \left( \ldots \right)    \right|   \\ 
&<&   | \beta  U_\beta |     +  |  \alpha  U_\alpha |       \quad{\rm as~}|\tanh|<1    \\ 
&=&   \left| \frac{u_0U_\beta}{U_\beta^2-1} \right| + \left| \frac{u_0U_\alpha}{U_\alpha^2-1} \right|   \\ 
&=&   |u_0|\left(   \frac{U_\beta}{U_\beta^2-1}  -   \frac{U_\alpha}{U_\alpha^2-1}  \right)   
    \quad {\rm as~}0<U_\alpha<1<U_\beta   \\    
&=&   |u_0|\left(   \frac{1}{U_\beta^2-1}  +  \frac{1}{U_\beta+1}  - \frac{1}{U_\alpha^2-1} -  \frac{1}{U_\alpha+1}    \right)  \\  
&<&   |u_0|\left(   \frac{1}{U_\beta^2-1} - \frac{1}{U_\alpha^2-1}  \right) \quad   {\rm as~} \frac{1}{U_\alpha+1}  > \frac{1}{U_\beta+1}  \\  
&=&   |\beta - \alpha| \ . 
\end{eqnarray*} 
\item   In this case we have 
\begin{eqnarray*}
|s_\alpha| + |\beta-\alpha| 
&=& |\alpha U_\alpha \tanh\left( \ldots \right) | + |\beta - \alpha | \\  
&<& |\alpha U_\alpha | + |\beta - \alpha |       \quad{\rm as~}|\tanh|<1    \\ 
&=&  \left| \frac{u_0U_\alpha}{U_\alpha^2-1} \right| + \left| \frac{u_0}{U_\beta^2-1} - \frac{u_0}{U_\alpha^2-1}  \right|   \\ 
&=& |u_0|\left(  \frac{U_\alpha}{1-U_\alpha^2}  + \frac{1}{1-U_\beta^2}  -   \frac{1}{1-U_\alpha^2}    \right)    
\quad   {\rm as~}  0<U_\alpha<U_\beta<1  \\
&=&  |u_0|\left( \frac{1}{1-U_\beta^2}  -  \frac{1}{1+U_\alpha}  \right)     \\ 
&=&  |u_0|\left( \frac{U_\beta}{1-U_\beta^2} + \frac1{1+U_\beta}  -  \frac{1}{1+U_\alpha}   \right)     \\
&<&  |u_0| \frac{U_\beta}{1-U_\beta^2}   \quad {\rm as~}  \frac1{1+U_\beta} <  \frac{1}{1+U_\alpha}     \\
&=&  |\beta U_\beta|   \\ 
&<&  |\beta U_\beta \coth\left( \ldots  \right) | \quad {\rm as~} |\coth| > 1   \\ 
&=& |s_\beta| \ . 
\end{eqnarray*} 
This contradicts $|s_\beta-s_\alpha|= |\beta-\alpha|$, as the latter implies  $|s_\beta| \le |s_\alpha| + |\beta-\alpha|$.  
\item Similarly to case 2 we have 
\begin{eqnarray*}
|s_\alpha| + |\beta-\alpha| 
&=& |\alpha U_\alpha \tanh\left( \ldots \right) | + |\beta - \alpha | \\  
&<& |\alpha U_\alpha | + |\beta - \alpha |       \quad{\rm as~}|\tanh|<1    \\ 
&=&  \left| \frac{u_0U_\alpha}{U_\alpha^2-1} \right| + \left| \frac{u_0}{U_\beta^2-1} - \frac{u_0}{U_\alpha^2-1}  \right|   \\ 
&=& |u_0|\left(  \frac{U_\alpha}{U_\alpha^2-1}  + \frac{1}{U_\beta^2-1}  -   \frac{1}{U_\alpha^2-1}    \right)    
\quad   {\rm as~}  1<U_\beta<U_\alpha    \\
&=&  |u_0|\left( \frac{1}{U_\beta^2-1}  +  \frac{1}{U_\alpha+1}  \right)     \\ 
&=&  |u_0|\left( \frac{U_\beta}{U_\beta^2-1} -  \frac1{U_\beta+1}  + \frac{1}{U_\alpha+1}   \right)     \\
&<&  |u_0| \frac{U_\beta}{U_\beta^2-1}   \quad {\rm as~}  \frac1{U_\alpha+1} <  \frac{1}{U_\beta+1}     \\
&=&  |\beta U_\beta|   \\ 
&<&  |\beta U_\beta \coth\left( \ldots  \right) | \quad {\rm as} |\coth| > 1   \\ 
&=& |s_\beta| \ . \quad 
\end{eqnarray*} 
\end{enumerate}

It remains to present the plots of some superpositions. 
Figure 9 shows a tanh-tanh superposition with $u_0=1$, $x_{\alpha,0}=0$, $x_{\beta,0}=2$, $c_\alpha=u_0-2\alpha=4$, 
$c_\beta=u_0-2\beta=-1$.
(For $u_0>0$ such solutions exist provded $c_\alpha> 3u_0$ and $c_\beta<u_0$ --- note here that $c_\beta$ 
can be positive or negative.)  
Figure 10 shows a tanh-coth soliton-soliton  superposition with $u_0=1$, $x_{\alpha,0}=0$, $x_{\beta,0}=-10$, $c_\alpha=u_0-2\alpha=4$, 
$c_\beta=u_0-2\beta=6$.
Figure 11 shows a tanh-coth cuspon-cuspon  superposition with $u_0=1$, $x_{\alpha,0}=0$, $x_{\beta,0}=-2$, $c_\alpha=u_0-2\alpha=-1$, 
$c_\beta=u_0-2\beta=-2$.

\begin{figure}[ht]
\begin{center}
\includegraphics[width=1.5in]{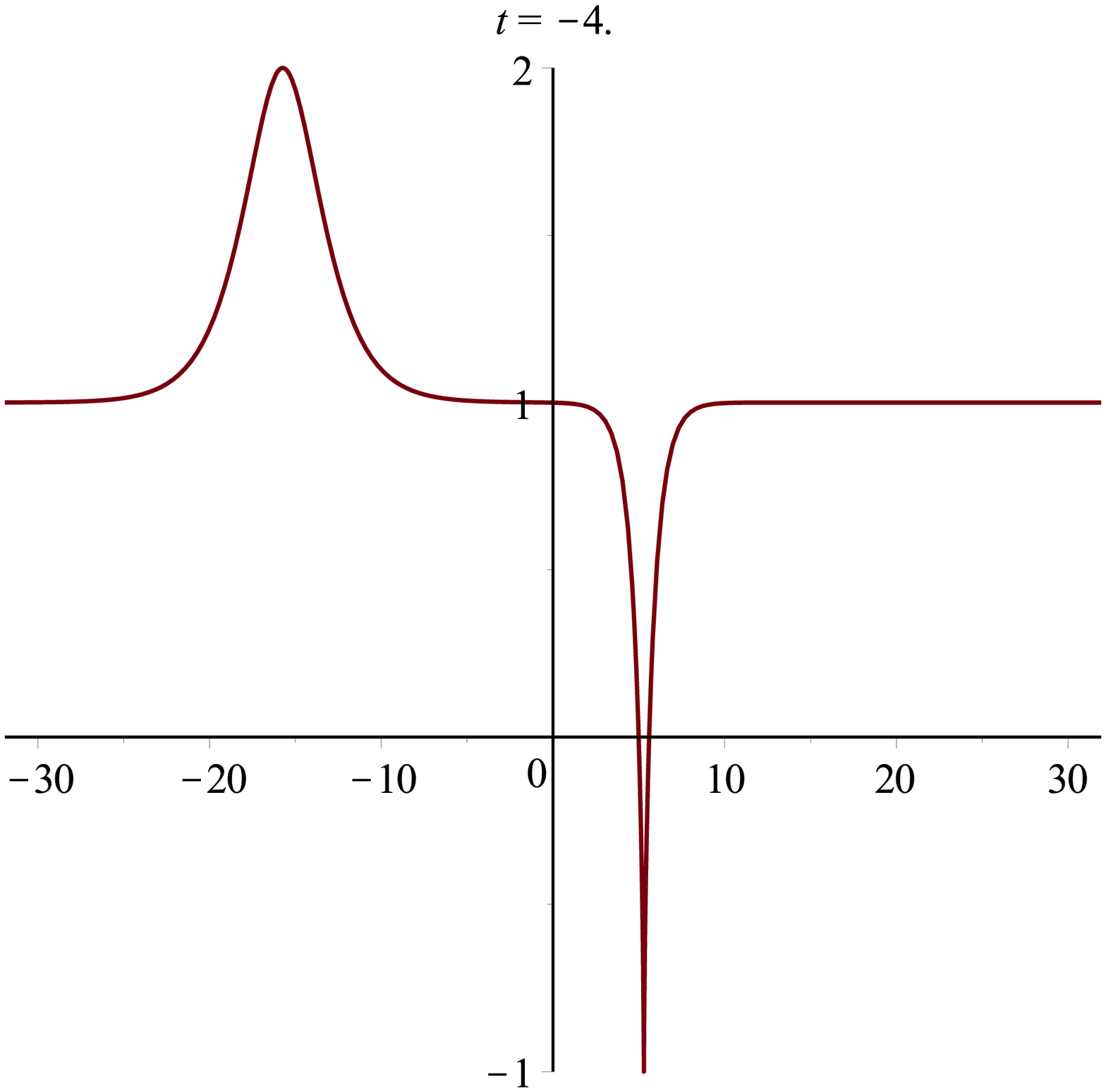}
\includegraphics[width=1.5in]{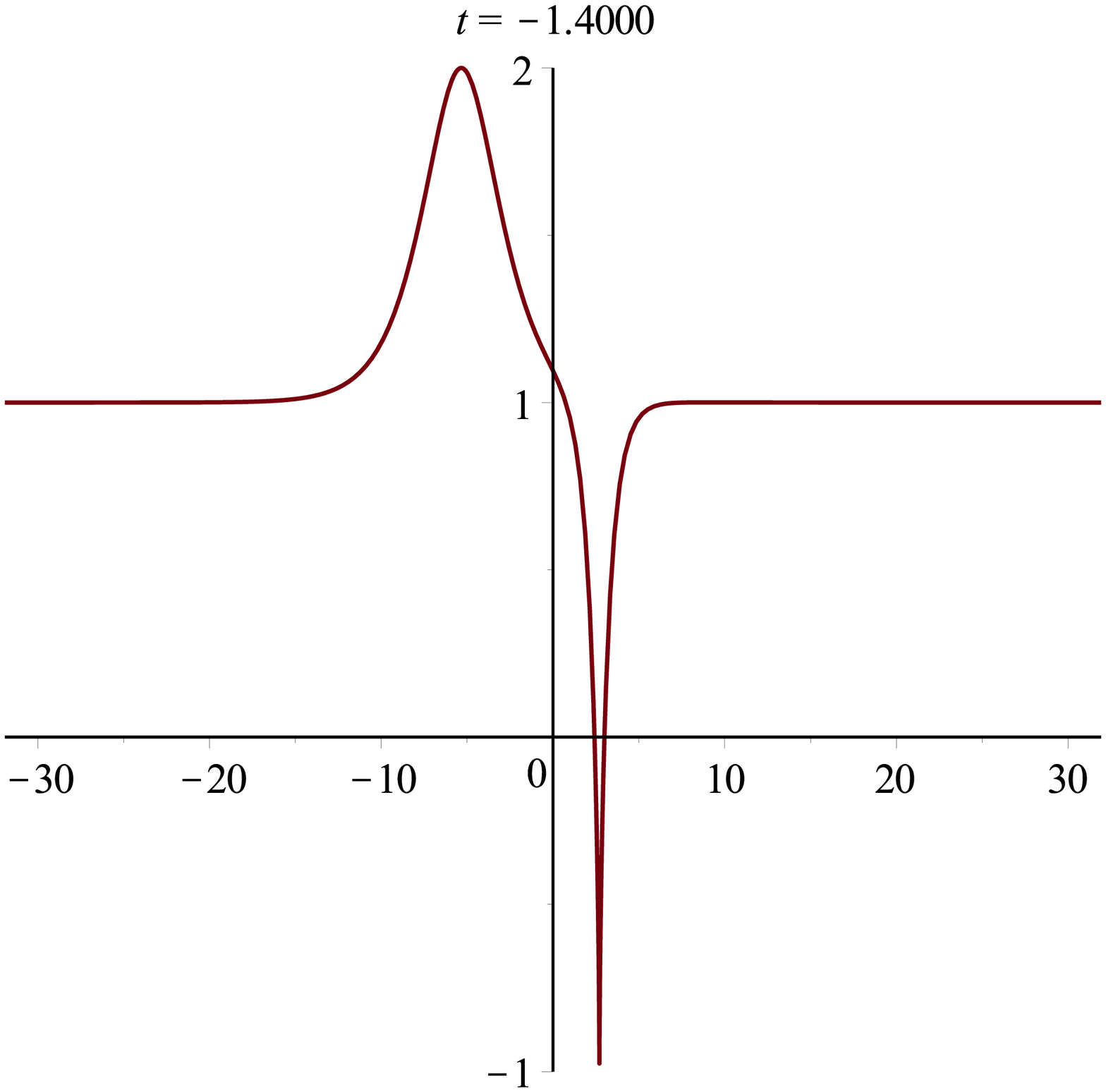}
\includegraphics[width=1.5in]{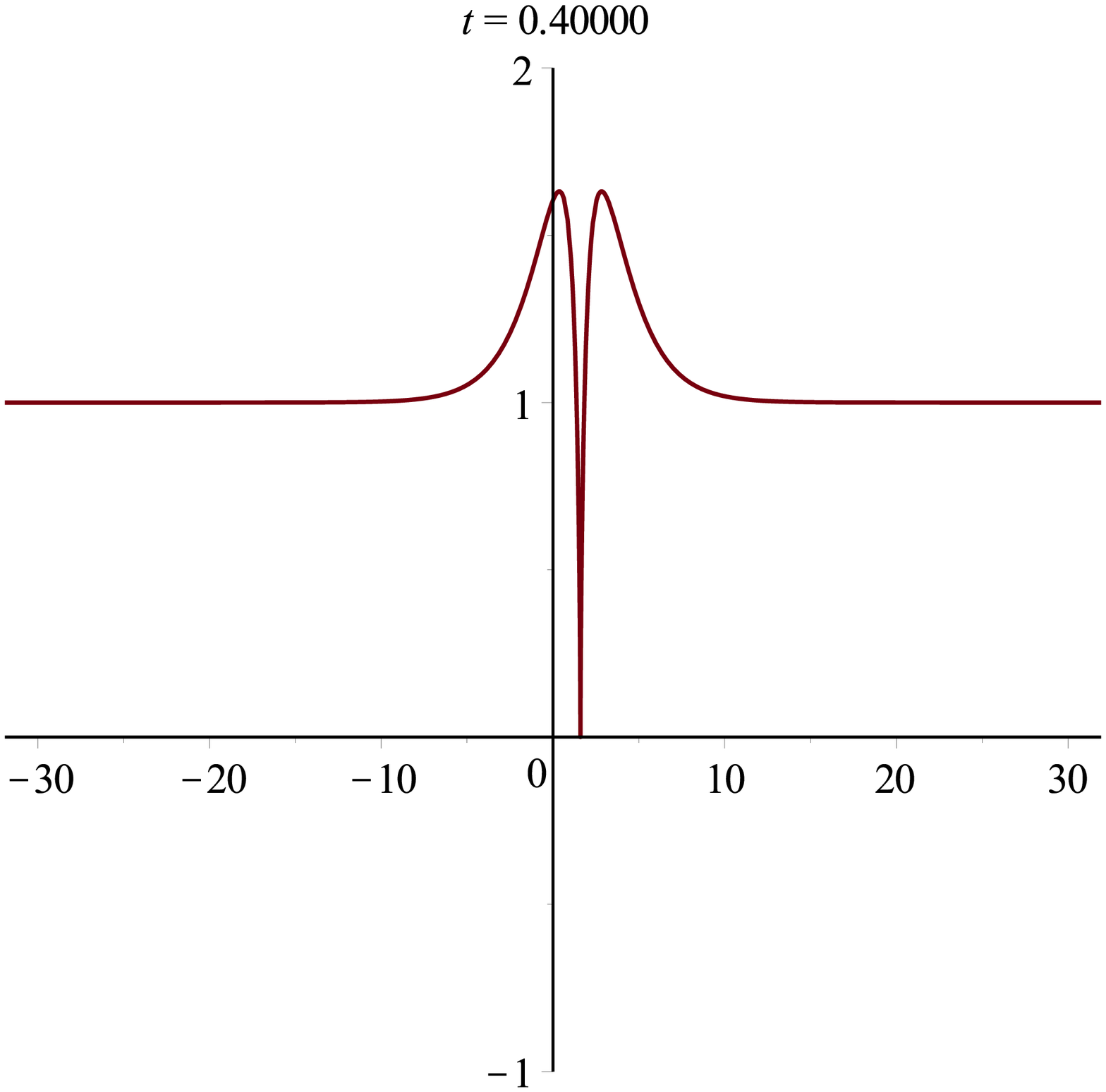}
\includegraphics[width=1.5in]{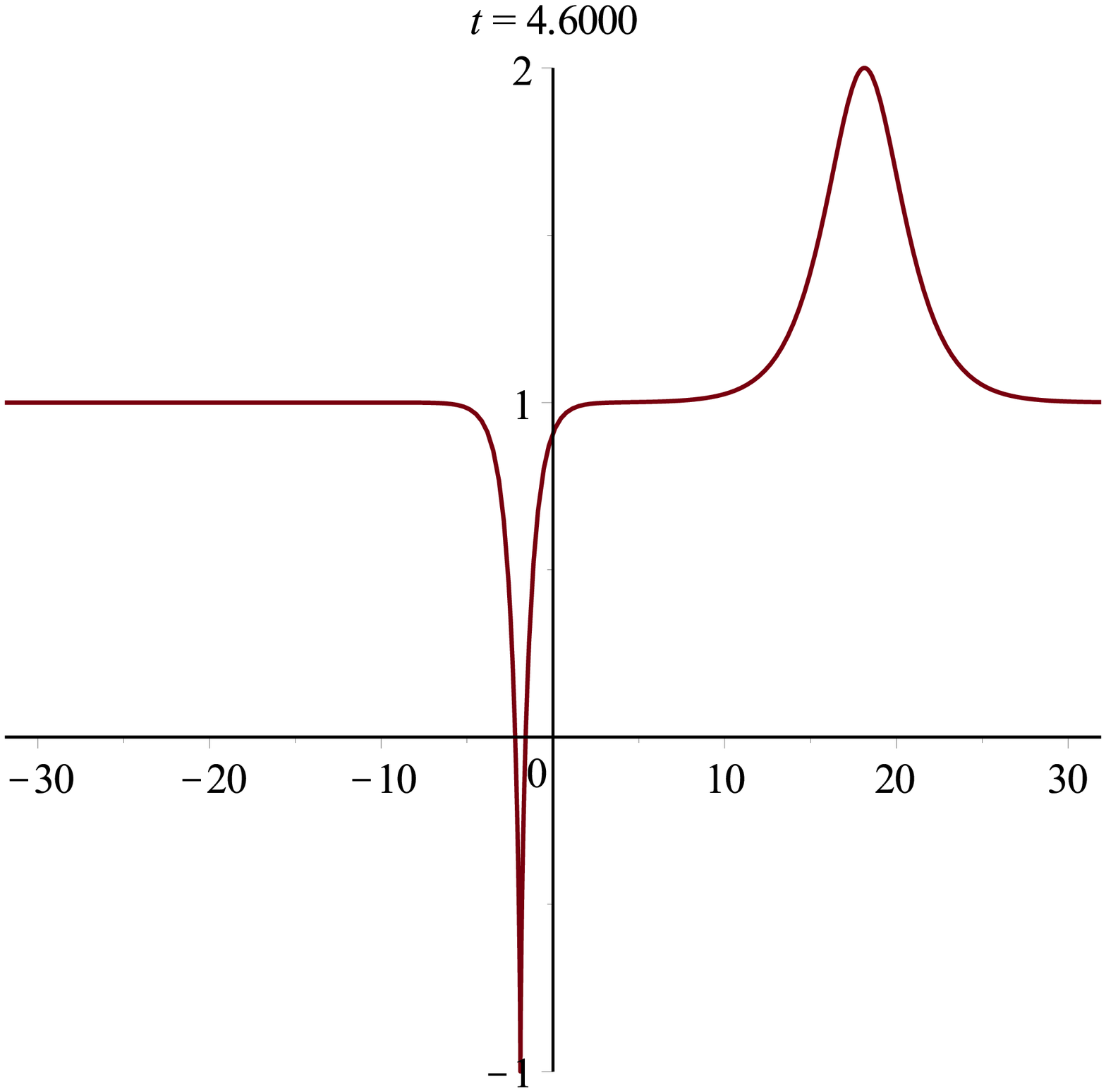}
\caption{Tanh-Tanh Superposition. $u_0=1$, $x_{\alpha,0}=0$, $x_{\beta,0}=2$, $c_\alpha=4$, $c_\beta=-1$.
Plots for times $t=-4,-1.4,0.4,4.6$ from left to right.}    
\end{center}
\end{figure}

\begin{figure}[ht]
\begin{center}
\includegraphics[width=1.5in]{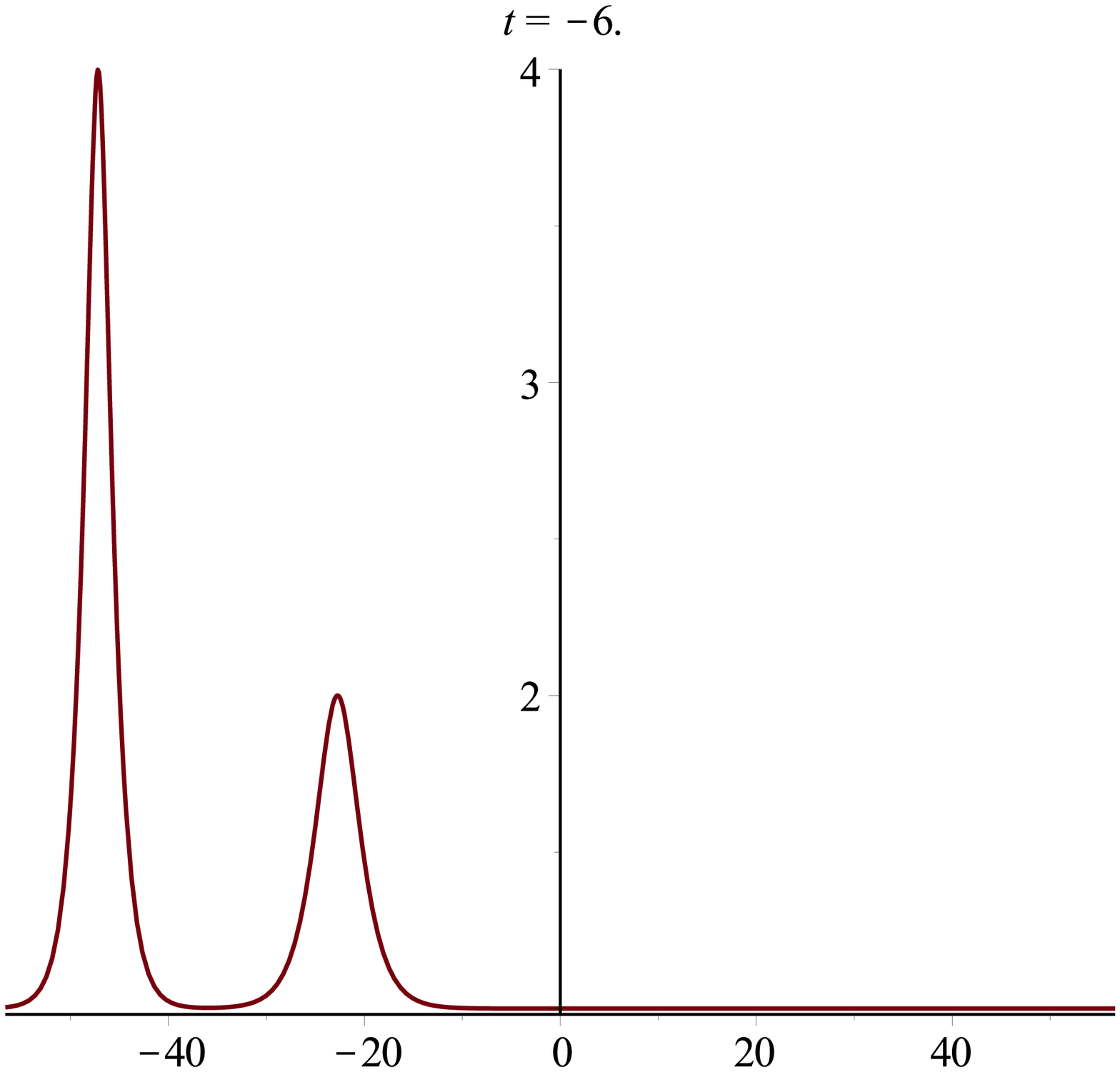}
\includegraphics[width=1.5in]{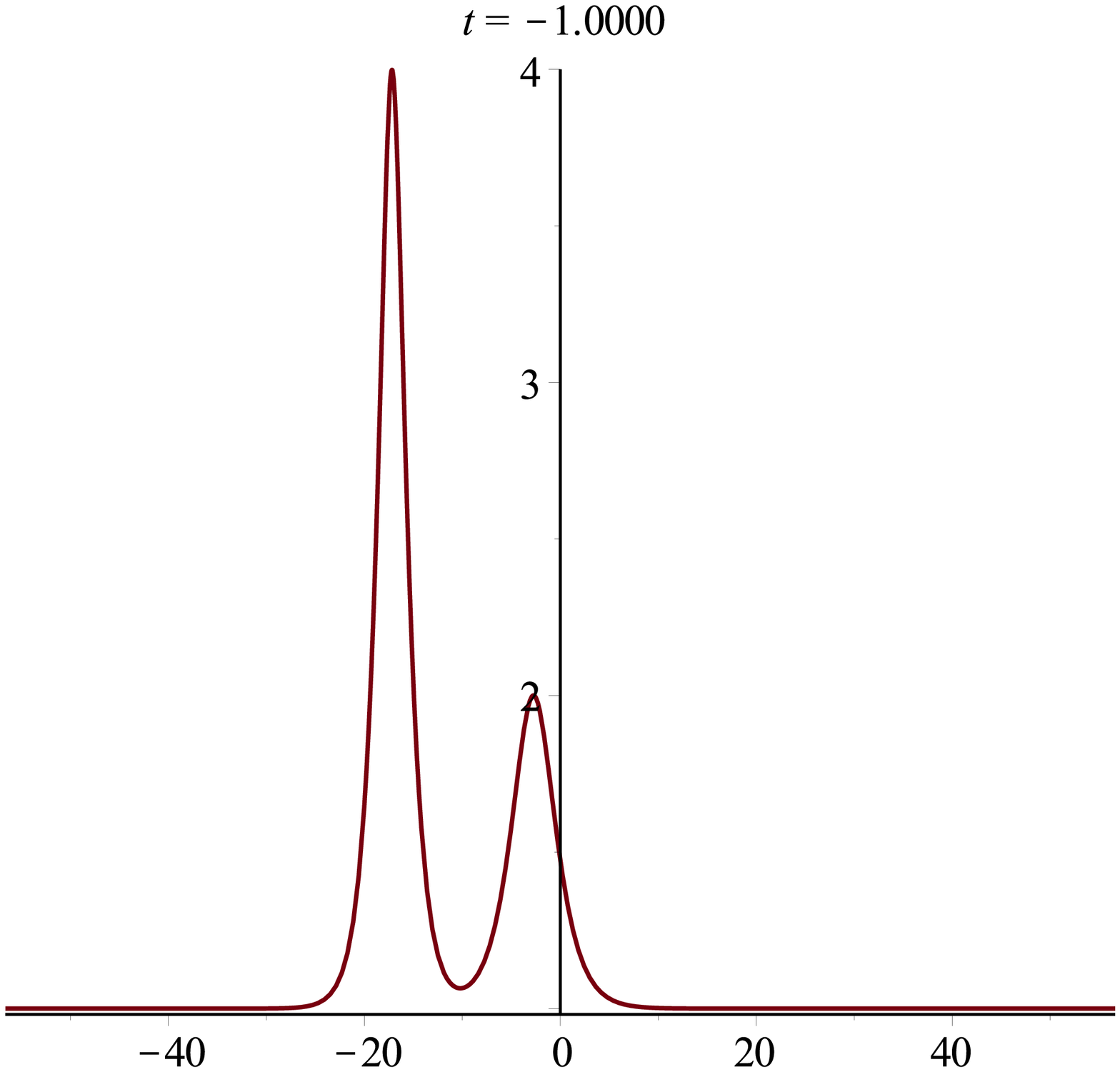}
\includegraphics[width=1.5in]{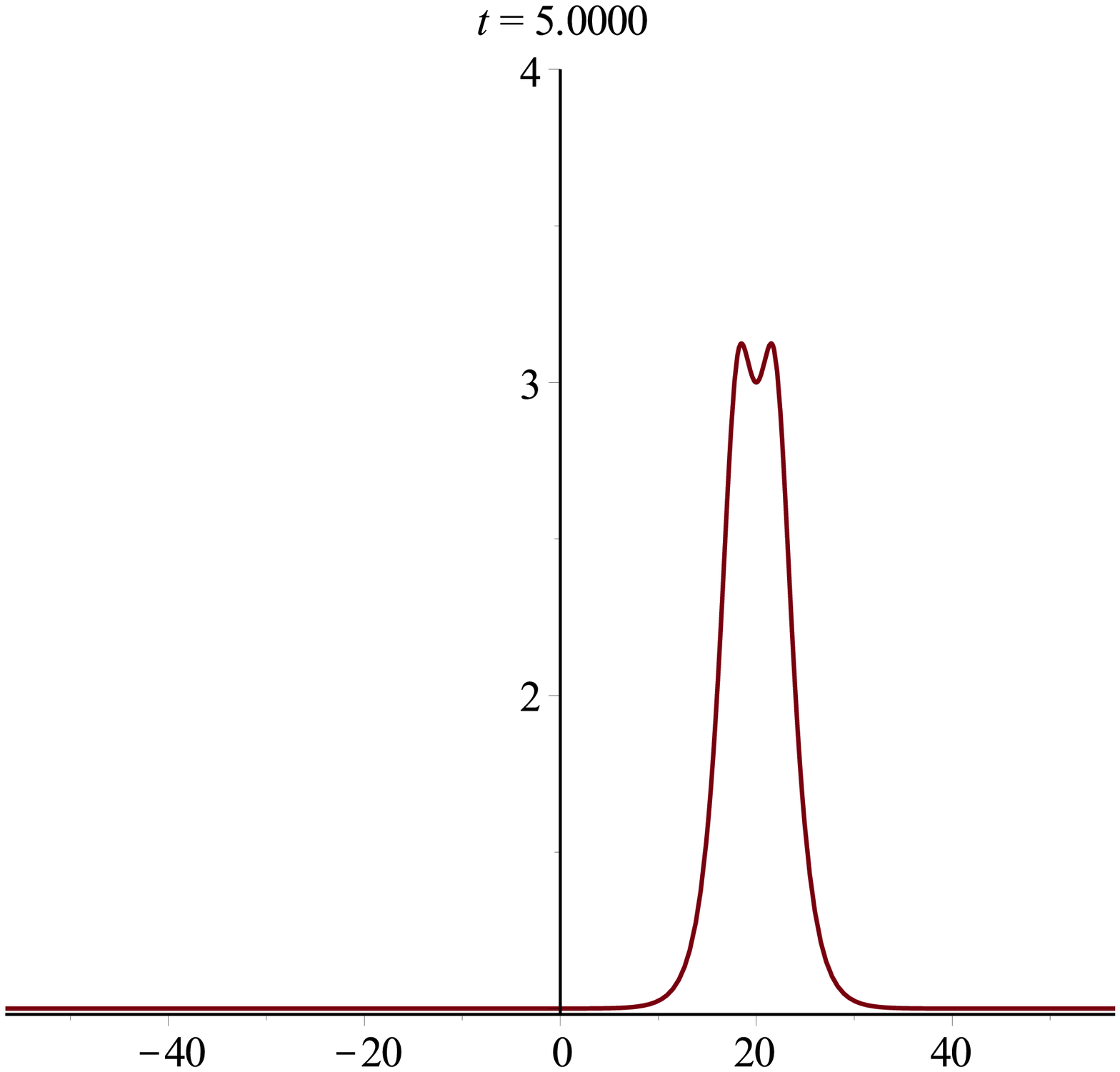}
\includegraphics[width=1.5in]{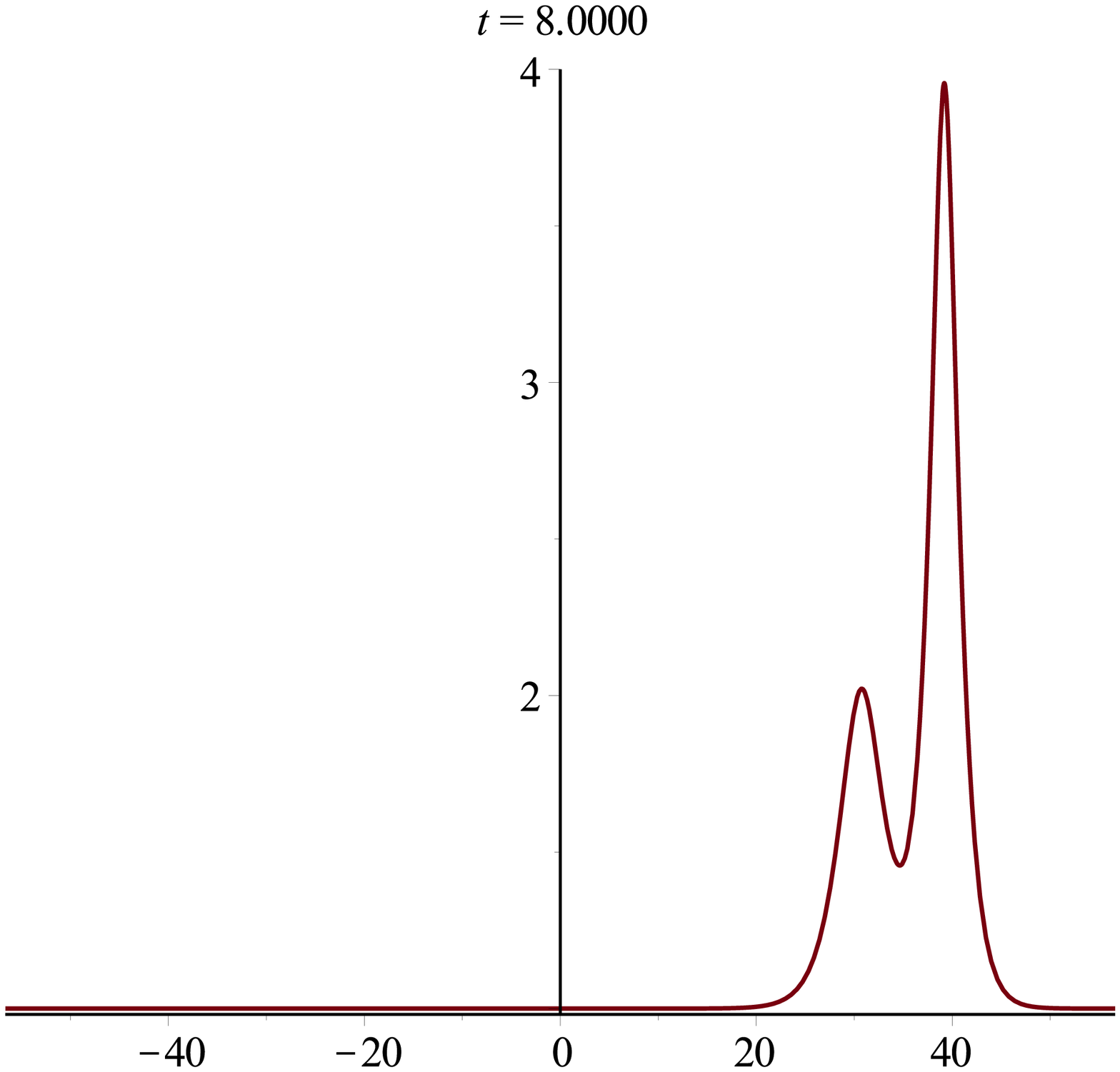}
\caption{Tanh-Coth Soliton-Soliton Superposition. $u_0=1$, $x_{\alpha,0}=0$, $x_{\beta,0}=-10$, $c_\alpha=4$, $c_\beta=6$.
Plots for times $t=-6,-1,5,8$ from left to right.} 
\end{center}
\end{figure}

\begin{figure}[ht]
\begin{center}
\includegraphics[width=1.5in]{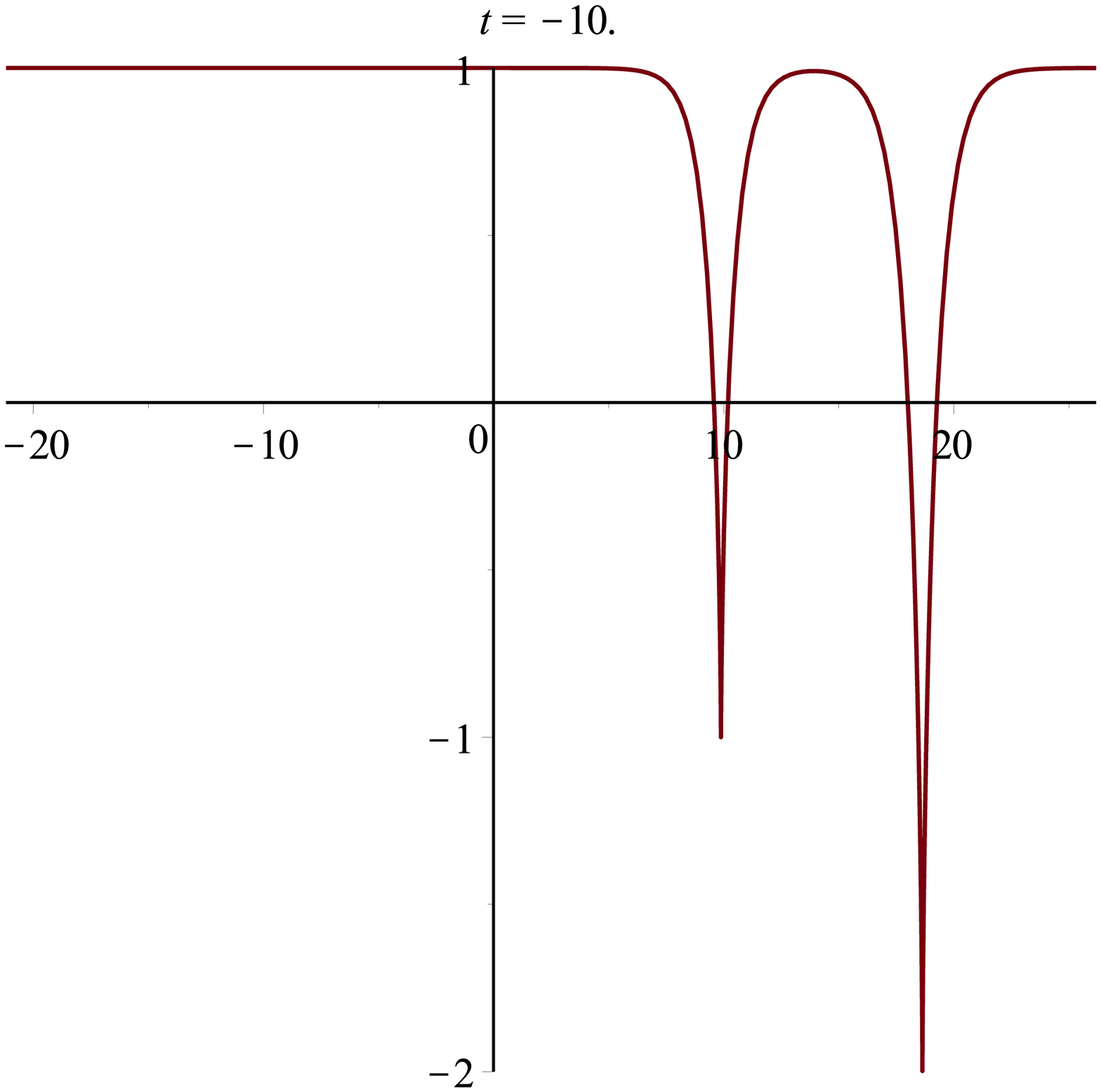}
\includegraphics[width=1.5in]{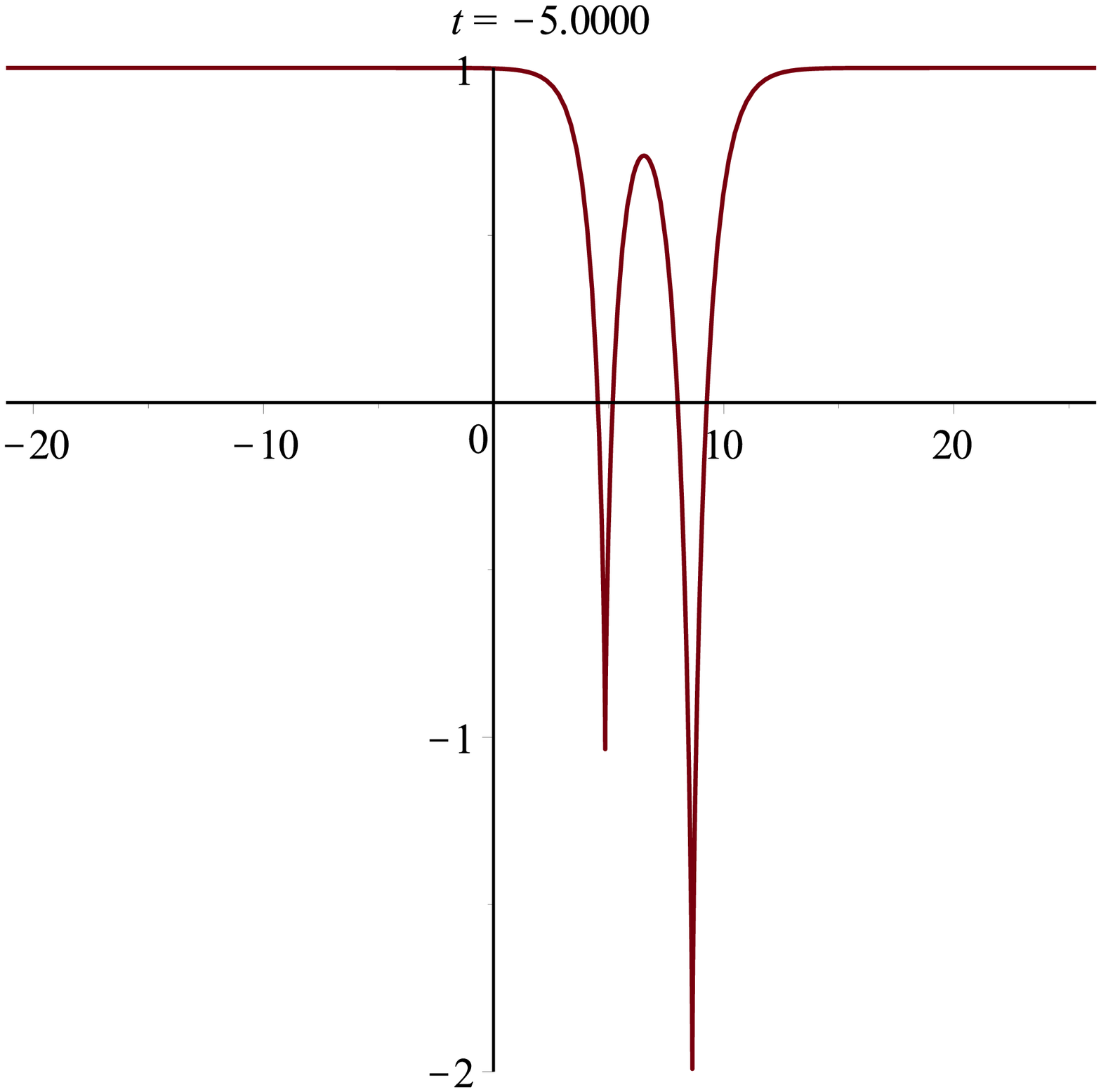}
\includegraphics[width=1.5in]{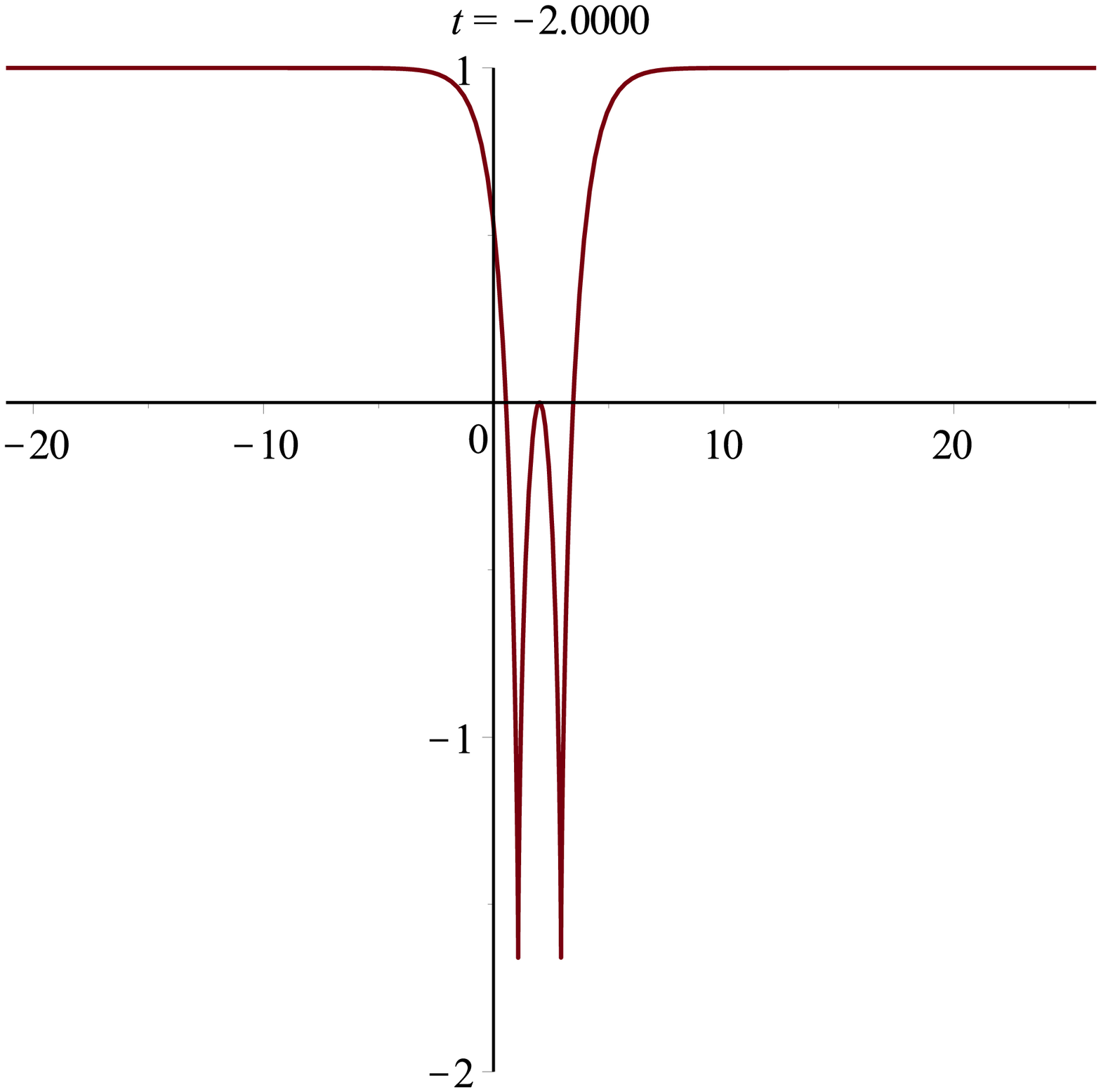}
\includegraphics[width=1.5in]{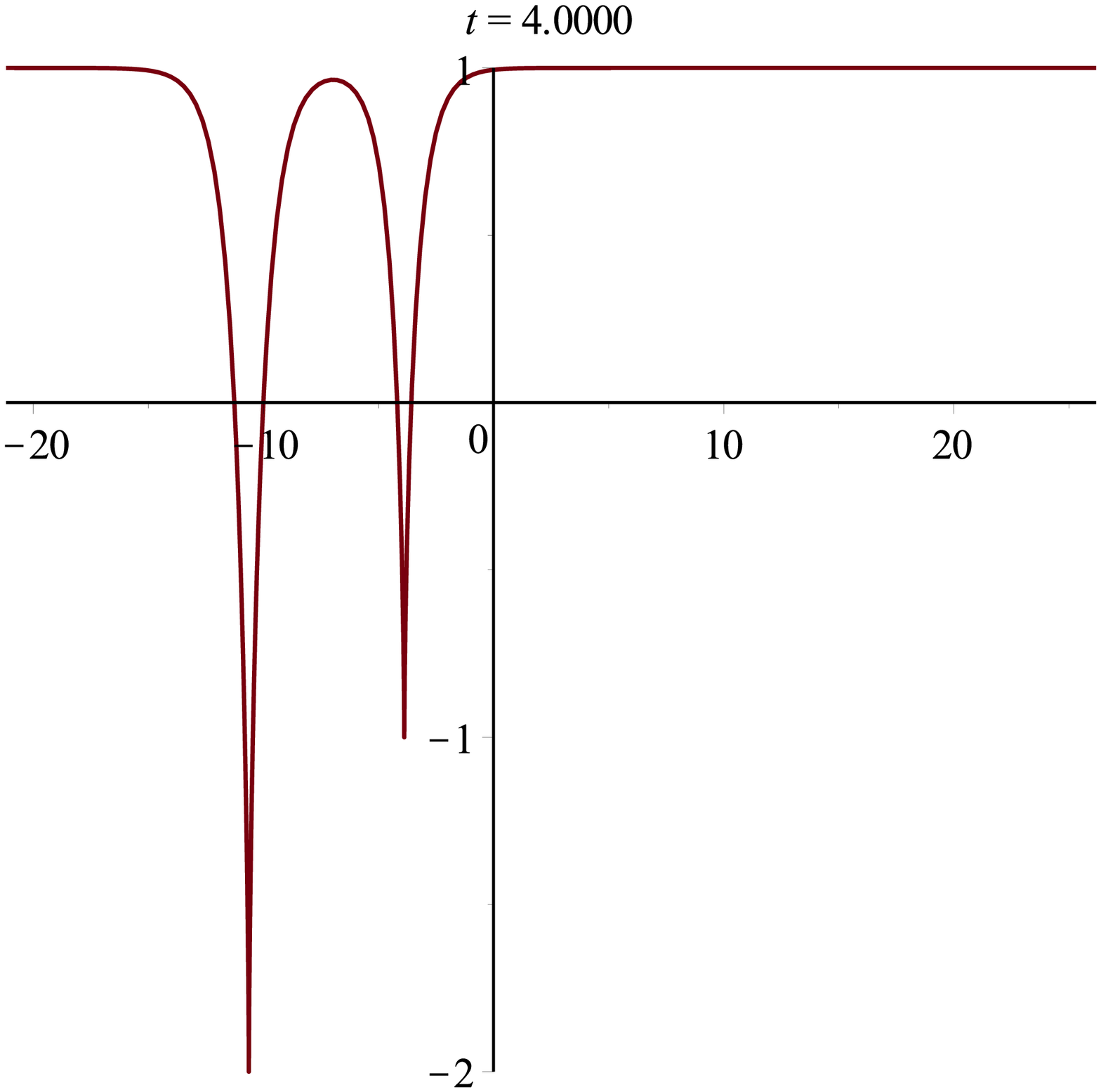}
\caption{Tanh-Coth Cuspon-Cuspon Superposition. $u_0=1$, $x_{\alpha,0}=0$, $x_{\beta,0}=-2$, $c_\alpha=-1$, $c_\beta=-2$.
Plots for times $t=-10,-5 , -2, 4$ from left to right.} 
\end{center}
\end{figure}

Note that in all the plots we have taken $u_0>0$, in which case, as in the previous section, 
the soliton solutions have positive speed $c>3u_0$ and central elevation $c-2u_0>u_0$, whereas 
the cuspon solutions have speed $c<u_0$, which can be positive or negative, and central elevation 
$c<u_0$ (i.e. they might be called ``anticuspons''). In the case $u_0<0$ everything is inverted and
reversed (reflecting the $t\rightarrow -t$, $u\rightarrow -u$ symmetry of (\ref{CM1}): solitons have 
negative speed $c<3u_0$ and central elevation $c-2u_0<u_0$ (antisolitons), and cuspons have 
speed $c>u_0$ and central elevation $c>u_0$. 

\section{Symmetries and conservation laws for the Camassa-Holm equation}

In this section we show how to use the BT to obtain infinite hierarchies  of symmetries 
and conservation laws for CH and pCH, following the general methodology described in \cite{RS2}. 
The discussion of symmetries of CH in the literature is limited, though the existence of an 
infinite number of symmetries is implicit from the bihamiltonian structure given in \cite{CH0}. 
In the series of papers \cite{rey0,rey1,HR1,reyP,HR2}, Reyes and collaborators present {\em nonlocal} symmetries 
of CH depending on a parameter, and then expand in powers of the parameter to obtain local symmetries, though 
limited details are given. Some explicit formulae appear  in \cite{GKKV}. Our approach is related, but we will not 
discuss the connection explicitly. 

As a starting point for our discussion of symmetries we could take 
the generating symmetry (\ref{QQ1}) for aCH, and work out the induced 
action on  $x$, the independent variable in CH. 
But a more direct approach is 
to look at the superposition principle (\ref{spu}),(\ref{spx}) in the limit that $\beta$ tends to $\alpha$,
but $s_\beta$ tends to a second solution of (\ref{BT1})-(\ref{BT2}) {\em distinct from}  $s_\alpha$. More explicity, 
setting $\beta=\alpha-\frac12 \epsilon$, $s_\alpha=s_\alpha^{(1)}$, $s_\beta=s_\alpha^{(2)}+O(\epsilon)$ in (\ref{spu}),(\ref{spx}) 
we obtain 
\begin{eqnarray*}
u_{\alpha,\alpha-\frac12 \epsilon} &=& u+\epsilon \frac{s_{\alpha}^{(1)}+s_{\alpha}^{(2)}+u_{x}}{s_{\alpha}^{(2)}-s_{\alpha}^{(1)}} +O(\epsilon^2)\ ,  \\
x_{\alpha,\alpha-\frac12 \epsilon} &=& x+\frac{\epsilon } {s_{\alpha}^{(2)}-s_{\alpha}^{(1)}}  + O(\epsilon^2)\ .  
\end{eqnarray*}
We deduce the generating symmetry for CH 
$X=Q^x\frac{\partial}{\partial x}+Q^u\frac{\partial}{\partial u}$ where
\begin{equation}
Q^x=\frac{1}{s_{\alpha}^{(2)}-s_{\alpha}^{(1)}}\ ,
\qquad 
Q^u=\frac{s_{\alpha}^{(1)}+s_{\alpha}^{(2)}+u_{x}}{s_{\alpha}^{(2)}-s_{\alpha}^{(1)}}\ .
\end{equation}
Here $s_{\alpha}^{(1)},s_{\alpha}^{(2)}$ are two different solutions of (\ref{BT1}),(\ref{BT2}) for the same parameter $\alpha$.
This symmetry depends upon $\alpha$; expansion in a (formal) power series in $\alpha$ will give an infinite hierarchy of symmetries.
However before we do this, we exploit the fact that a generalized symmetry of the form 
$X=Q^x\frac{\partial}{\partial x}+Q^u\frac{\partial}{\partial u}$ which acts on both the dependent and independent variables can 
be transformed to a generalized symmetry which acts only on the dependent variable \cite{ol0}
with characteristic $Q=Q^u-Q^xu_x$. Here we have 
\begin{equation}
Q=\frac{s_{\alpha}^{(1)}+s_{\alpha}^{(2)}}{s_{\alpha}^{(2)}-s_{\alpha}^{(1)}}\ .\label{symm1}
\end{equation}
This is also the characteristic for a  symmetry of the full family of equations  (\ref{GCH}).

The next thing to do is to find a (formal) asymptotic series solution of (\ref{BT1})-(\ref{BT2}) for small $|\alpha|$. 
This takes the form
\begin{equation}
s_{\alpha}=\sum_{n=1}^{\infty}s_n\alpha^{\frac{n}{2}},\label{exp1}
\end{equation}
where
$$
s_1=\sqrt{m},~~~s_2=-\frac{s_{1,x}}{s_1},~~~s_{n+1}=-\frac{s_{n,x}}{s_1}+\frac{1}{2s_1}\left(\delta_{n,2}-\sum_{i=0}^{n-2}s_{i+2}s_{n-i}\right),~~~n=2,3... .
$$
A second solution of (\ref{BT1})-(\ref{BT2}) can be obtained by replacing $\alpha^{\frac{1}{2}}$ by $-\alpha^{\frac{1}{2}}$. So we get
\begin{equation}
s_{\alpha}^{(1)}=\sum_{n=1}^{\infty}s_n\alpha^{\frac{n}{2}},~~~s_{\alpha}^{(2)}=\sum_{n=1}^{\infty}s_n(-\alpha^{\frac{1}{2}})^n.
\label{exp2} 
\end{equation}
Plugging this into (\ref{symm1}) we obtain
\begin{equation}
\frac{Q}{\sqrt{\alpha}}=\frac{\sum_{n=1}^{\infty}s_{2n}\alpha^{n}}{\sum_{n=1}^{\infty}s_{2n-1}\alpha^{n}}.\label{csch}
\end{equation}
The expansion of (\ref{csch}) around $\alpha=0$ gives an infinite hierarchy of symmetries of CH.
The first few of these take the form
\begin{eqnarray}
X_{1}&=&\left(\frac{1}{\sqrt{m}}\right)_x\frac{\partial}{\partial u}\ ,  \\
X_{2}&=&\left(\frac{4mm_{xx}-5m_x^2+4m^2}{m^{7/2}}\right)_x\frac{\partial}{\partial u}\ ,  \\
X_{3}&=&\left(\frac{
\begin{array}{cc} 
 64 m^3  m_{xxxx} -448 m^2   m_x m_{xxx} -160 m^3 m_{xx} +1848 m  m_x^2 m_{xx}   \\
-336 m^2  m_{xx}^2 +280 m^2  m_x^2  -1155  m_x^4 -48 m^4
\end{array}
}{m^{13/2}}\right)_x\frac{\partial}{\partial u}\ .   
\end{eqnarray}
The fact that all the characteristics are $x$-derivatives is indicative that these symmetries can be derived 
from corresponding symmetries of pCH. The generating symmetry for pCH (up to an irrelevant overall constant factor) is 
thus  $Q^v  \frac{\partial}{\partial v}  $ where 
\begin{equation}
Q^{v} = \frac{1}{s_{\alpha}^{(1)}-s_{\alpha}^{(2)}}  . \label{Qv}
\end{equation}

As stressed before, the symmetry with characteristic (\ref{symm1}) is a symmetry for the full family of equations 
(\ref{GCH}), including the HS equation. 
For HS the asymptotic series solutions of (\ref{BTHS1})-(\ref{BTHS2}) takes the form
\begin{equation}
s^{(1)}_{\alpha}=\sum_{n=1}^{\infty}s_n\alpha^{\frac{n}{2}}\ ,\qquad  s^{(2)}_{\alpha}=\sum_{n=1}^{\infty}s_n(-\alpha^{\frac{1}{2}})^n \ , 
\end{equation}
where
$$
s_1=i\sqrt{u_{xx}},~~~s_2=-\frac{s_{1,x}}{s_1},~~~s_{n+1}=-\frac{s_{n,x}}{s_1}-\frac{1}{2s_1}\left(\sum_{i=0}^{n-2}s_{i+2}s_{n-i}\right),~~~n=2,3... .
$$
Proceeding as before gives an infinite hierarchy of symmetries for HS, with the first few taking the form
\begin{eqnarray*}
X_{1}&=&\left(\frac{1}{\sqrt{u_{xx}}}\right)_x\frac{\partial}{\partial u}\ ,\\
X_{2}&=&\left(4\frac{u_{xxxx}}{u_{xx}^{5/2}}-5\frac{u_{xxx}^2}{u_{xx}^{7/2}}\right)_x\frac{\partial}{\partial u}\ ,\\
X_{3}&=&\left(1155\,{\frac {u_{xxx}^{4}}{u_{xx}^{13/2}}}-1848\,{\frac {u
_{xxx}^{2}u_{xxxx}}{u_{xx}^{11/2}}}+448\,{\frac {u_{xxx}u_{xxxxx}}{u_{xx}^{9/2}}}
+336\,{\frac {u_{xxxx}^{2}}{u_{xx}^{9/2}}}-64\,{\frac {u_{xxxxxx}}{u_{xx}^{7/2}}}\right)_x\frac{\partial}{\partial u}\ .
\end{eqnarray*}

Using the fact that if a single 
solution of the Riccati equation (\ref{BT1}) is known then it is possible to find
the general solution by quadratures, it is possible to rewrite (\ref{symm1}) in the form 
$$
Q    = 1 +   \frac{s_{\alpha}^{(1)}(x)}{\alpha} 
     \left(  \int_{x_0}^x  e^{\frac{1}{\alpha}\int_y^x s_{\alpha}^{(1)}(z)dz}  dy + C e^{\frac{1}{\alpha}\int_{x_0}^x s_{\alpha}^{(1)}(z)dz}  \right)  \ . 
$$
and (\ref{Qv}) in the form 
$$
Q^v =  \int_{x_0}^x  e^{\frac{1}{\alpha}\int_y^x s_{\alpha}^{(1)}(z)dz}  dy + C e^{\frac{1}{\alpha}\int_{x_0}^x s_{\alpha}^{(1)}(z)dz} \ . 
$$
Here $C$ is an arbitrary constant. Since a linear combination of symmetries is a 
symmetry, both terms  on the RHS are by themselves the characteristics of  symmetries. 
The symmetry associated with the factor multiplying $C$ is the nonlocal symmetry first presented in \cite{rey0}. 
The relation between B\"acklund transformations and nonlocal symmetries has recently been discussed in \cite{LHC}.

A conservation law (CL) for a PDE for the scalar function $u(x,t)$  is an expression
$$
T_t+X_x=0 
$$
which holds  on solutions of the equation. 
Conservation laws for CH can be obtained from (\ref{BTT}) by writing it in the form
\begin{equation}
s_t+\left(su-\alpha(u_x+2s)\right)_x=0   \label{precons} 
\end{equation}
Using the expansion (\ref{exp1}) for $s$ in (\ref{precons}) we obtain an infinite hierarchy of conservation laws. 
Terms with integer powers of $\alpha$ in this expansion give trivial CLs. To prove this, observe 
from (\ref{exp2}) that the terms with integer powers are obtained by setting 
$s=\frac12 \left( s_\alpha^{(1)}+ s_\alpha^{(2)} \right)$ in (\ref{precons}). But from (\ref{BT1}) it is 
simple to verify that 
$$  \frac12 \left( s_\alpha^{(1)}+ s_\alpha^{(2)} \right) =   -\alpha \left( \ln 
| s_\alpha^{(1)}- s_\alpha^{(2)} | \right)_x \ . 
$$
Thus to obtain nontrivial laws we look at only the half-integer powers of $\alpha$. Thus we set 
$s=s_\alpha^{(1)}- s_\alpha^{(2)}$ in (\ref{precons}) to obtain the generating conservation law 
\begin{eqnarray}
T &=& s_{\alpha}^{(1)}-s_{\alpha}^{(2)} \ ,  \\ 
X &=& (u-2\alpha)(s_{\alpha}^{(1)}-s_{\alpha}^{(2)}) \ .
\end{eqnarray}
The expansion of the generating conservation law around $\alpha=0$ gives an 
infinite hierarchy of nontrivial CLs for CH and pCH. The first few take the form
\begin{eqnarray*}
T_1 &=&  \sqrt {m}  \ ,  \\
X_1 &=& uT_1 \ , \\
T_2 &=& \frac{1}{8m^{5/2}}\left(4m^2+4mm_{xx} -5m_x^2\right) \ ,  \\
X_2 &=& uT_2-2T_1  \ ,  \\
T_3 &=& \frac{1}{128m^{11/2}} \left( 
64 m^3 m_{xxxx} - 1105m_x^4  + 1768mm_x^2m_{xx} -304 m^2 m_{xx}^2 - 448 m^2m_xm_{xxx} \right. \\
  && \left. -96 m^3m_{xx} + 200 m^2 m_x^2 -16m^4  \right) \ , \\
X_3 &=& uT_3-2T_2  \ . 
\end{eqnarray*}
Similar results can be obtained for HS and the full family of equations (\ref{GCH}). The existence of an infinite number 
of conservation laws for CH follows from the bihamiltonian structure for CH discovered in \cite{CH0}. The 
local form of these conservations laws was first obtained in \cite{FS1}, and they were subsequently further 
studied, and their derivation simplified, in numerous works such as 
\cite{rey0,rey1,HR1,reyP,HR2,Len2,CLOP,Iv1,GKKV}. The derivation given here can be easily related to previous ones, 
though the use of 2 solutions of (\ref{BT1})-(\ref{BT2}) to understand the triviality of ``half'' of the conservation 
laws, is, we believe, new. 

\section{Concluding remarks}
In this paper we have explored the theory of the B\"acklund transformation for the Camassa-Holm 
equation. This is an unfamiliar type of BT, as it acts on one of the independent variables, as well as the 
dependent variables. However, it has emerged that it is just as useful --- using the superposition principles 
for the action on the different variables, we can exploit the BT to write down two wave solutions, just as 
is done for standard integrable equations such as KdV. Furthermore, we have shown how a double BT encodes an infinite
set of symmetries for CH, and the relationship of the BT and conservation laws. 

We have seen that the BT can also generate ``unphysical'' solutions, by which we mean solutions for which the new 
independent variable is not a $1-1$ function of the old independent variable. Going beyond two wave solutions, it is not
clear exactly what superpositions are allowed without creating singularities, though it seems to be a reasonable hypothesis that 
all possible combinations of solitons and cuspons can be formed, with the speeds permitted by the value of $u_0$, as listed 
in the table at the end of section 5. It seems to us that this is a problem that remains to be handled indepdendent of the 
method used for constructing multiwave solutions. 

Peakons emerge from both solitons and cuspons in the limit $u_0\rightarrow 0$ (with one giving rise to peakons of positive 
speed and one to peakons of negative speed, depending on whether the limit is taken from below or from above). 
This is an extremely singular limit. We have not yet found a way to apply a superposition principle directly to peakons,
but we continue to search.  

Finally, one more general comment. The BT, in its minimalist form, is the transformations (\ref{BT4}) and (\ref{BT3}) 
where $s$ satisfies (\ref{BT1})-(\ref{BT2}). The latter equations for $s$ are equivalent to the Lax pair, or linear 
system, for CH. So the BT seems to be more than the linear system. We wonder if there is a case of an integrable system without
a BT? 

\bibliographystyle{acm} 
\bibliography{P} 
\end{document}